\documentclass[aps,prb,twocolumn,groupedaddress,showpacs,floatfix,altaffilletter, include graphics, longbibliography]{revtex4-2}
\usepackage{graphicx}
\usepackage{grffile}
\usepackage{amsmath}
\usepackage{amssymb}
\usepackage{bm}
\usepackage{color}
\usepackage[dvipsnames]{xcolor}
\usepackage{amsmath,amssymb,amsfonts}
\usepackage{epsfig}
\usepackage{times}
\usepackage[colorlinks,bookmarks=false,citecolor=blue,linkcolor=red,urlcolor=blue]{hyperref}
\usepackage{gensymb}
\usepackage[toc,page]{appendix}
\usepackage{float}

\newcommand{\fref}[1]{Fig.~\ref{#1}}
\newcommand{\eref}[1]{Eq.~(\ref{#1})} 
\newcommand{\sref}[1]{Sec.~(\ref{#1})}

\begin{document}

\title{An Improved Two-Particle Self-Consistent Approach}

\author{C. Gauvin-Ndiaye$^{1}$, C. Lahaie$^{1}$, Y.M. Vilk$^2$, and A.-M.S.~Tremblay$^{1}$}
\affiliation{$^1$D{\'e}partement de Physique, Institut quantique, and RQMP Universit{\'e} de Sherbrooke, Sherbrooke, Qu{\'e}bec, Canada  J1K 2R1 \\
$^2$ 33 Weatherly Dr, Salem, MA 01970}
\date{\today}
\begin{abstract}
The two-particle self-consistent approach (TPSC) is a method for the one-band Hubbard model that can be both numerically efficient and reliable. However, TPSC fails to yield physical results deep in the renormalized classical regime of the bidimensional Hubbard model where the spin correlation length becomes exponentially large. We address the limitations of TPSC with improved approaches that we call TPSC+ and TPSC+SFM. In this work, we show that these improved methods satisfy the Mermin-Wagner theorem and the Pauli principle. We also show that they are valid in the renormalized classical regime of the 2D Hubbard model, where they recover a generalized Stoner criterion at zero temperature in the antiferromagnetic phase. We discuss some limitations of the TPSC+ approach with regards to the violation of the f-sum rule and conservation laws, which are solved within the TPSC+SFM framework. Finally, we benchmark the TPSC+ and TPSC+SFM approaches for the one-band Hubbard model in two dimensions and show how they have an overall better agreement with available diagrammatic Monte Carlo results than the original TPSC approach.

\end{abstract}

\maketitle


\section{Introduction}
\label{sec:intro}
As one of the simplest models to encapsulate the effect of strong correlations in electronic systems, the Hubbard model has been used to study quantum materials such as the cuprates \cite{Qin_2022} and the newly discovered nickelate superconductors \cite{Kitatani_2020}. The accurate description of realistic systems often requires the use of extensions of the Hubbard model to the multi-orbital case. For instance, the cuprates seem to be described more accurately by the three-band VSA~\cite{Varma_Schmitt-Rink_Abrahams_1987} Emery-Hubbard model \cite{Emery_1987}, strontium ruthenate by a three-band Hund's metal with strong spin-orbit coupling \cite{Acharya_2017}, and the nickelates by a model that takes into account at least one correlated band and a charge reservoir \cite{Held_2022}. 

One of the challenges in studying strongly correlated electron systems is that even the simpler one-band Hubbard model has no exact analytical solution for dimensions other than $1$ \cite{Lieb_1968} and infinity \cite{Metzner_1989}. The numerical solution of the model in finite dimensions $d>1$ can be achieved through approximate methods such as dynamical mean-field theory (DMFT) \cite{Metzner_1989, Georges_1992, Georges_1996}, its cluster extensions \cite{Hettler_1998, Lichtenstein_2000, Kotliar_2001} and diagrammatic extensions \cite{Rohringer_2018}, or through numerically exact quantum or diagrammatic Monte Carlo simulations \cite{Blankenbecler_1981, Prokofev_1998, Rossi_2017}.
Many of the methods for the one-band Hubbard model have recently been reviewed and benchmarked extensively for the $2D$ weak-coupling case at half-filling \cite{Schafer_2021}. At low temperatures, multiple methods face challenges even in the weak-coupling regime. For instance, finite cluster sizes limit the use of determinantal quantum Monte Carlo (DQMC) and cluster extensions of DMFT when the correlation length becomes large, while diagrammatic Monte Carlo (DiagMC) becomes limited by convergence issues. Diagrammatic extensions of DMFT such as the dynamical vertex approximation (D$\Gamma$A) can be used in a wider range of low temperatures. However, such methods are computationally expensive, which limits their application to extended, multi-orbital Hubbard-like models. 

The two-particle self-consistent approximation (TPSC) is a conserving, non-perturbative method for the Hubbard model that can be derived from a Luttinger-Ward functional \cite{Vilk_1997,TremblayMancini:2011}. In this approximation, the double occupancy is calculated self-consistently from the local spin and charge sum rules. The TPSC approximation introduces RPA-like spin and charge susceptibilities with renormalized vertices $U_{sp}$ and $U_{ch}$. These susceptibilities and vertices are then used to compute the self-energy. 

The TPSC approximation is the first method that predicted the opening of an antiferromagnetic pseudogap in the $2D$ Hubbard model at weak coupling, and that quantitatively associated the phenomenon with the Vilk criterion \cite{Vilk_1997, Kyung_2004, Schafer_2021}. That criterion states that an antiferromagnetic pseudogap opens up when the spin correlation length exceeds the thermal de Broglie wave length. 

Though the TPSC approximation was first formulated in the context of the one-band Hubbard model, it has since been extended to the multi-orbital case \cite{Miyahara_2013, Zantout_2021}. The flexibility of this method, its low computational cost, as well as the fact that it respects the Mermin-Wagner theorem, the Pauli principle and conservation laws, make it attractive for applications to real materials and for extensions such as out-of-equilibrium calculations \cite{Simard_2022}.

However, the TPSC approximation has many limitations, such as the fact that it is only valid in the weak to intermediate interaction regime of the Hubbard model. It is also not valid deep in the renormalized classical regime in $2D$. Though it agrees quantitatively with benchmarks at high temperatures \cite{Vilk_1997,Moukouri:2000,TremblayMancini:2011} and can give a qualitative description of the crossover to the renormalized classical regime, it overestimates spin fluctuations at low temperatures, leading some of its predictions such as the self-energy and the double occupancy to deviate significantly from the benchmarks \cite{Schafer_2021}.

In this work, we introduce an improved version of TPSC, which we call TPSC+. In addition to better agreement with benchmarks, TPSC+ has the advantage that in two dimensions it is valid deep in the pseudogap regime, all the way to the zero-temperature long-range ordered antiferromagnet. Very few approximations~\cite{Borejsza_Dupuis_2003,Borejsza:2004,SangiovanniAFM:2006} can achieve this. The TPSC+ approximation, that includes some level of self-consistency, was first discussed in Ref. \cite{Schafer_2021}, but we provide here an extended discussion of its properties. In \sref{sec:model}, we discuss the model and obtain equations for the self-energy and generalized susceptibilities using the functional derivative approach. Next, we review the TPSC equations in \sref{sec:tpsc} and give an overview of its main properties. We introduce the TPSC+ approximation formalism in \sref{sec:tpsc+}. We also discuss the limitations of the method, more precisely how it violates spin and charge conservation laws and the f-sum rule. We show that a variant of the TPSC+ approximation, the TPSC+SFM method, can mitigate these limitations. Finally, we show the application of the TPSC+ and TPSC+SFM approximations to the $2D$ Hubbard model in \sref{sec:results}, where we provide comparisons to DiagMC \cite{Schafer_2021, Gukelberger_2015} and CDet \cite{Simkovic_2021} benchmarks. We show that the TPSC+ and the TPSC+SFM approximations are valid in the weak to intermediate regime of the $2D$ Hubbard model and that they outperform the original TPSC approximation at low temperatures, while maintaining low computational costs.


\section{Model and exact results}
\label{sec:model}
We start with the definition of the Hubbard model in \sref{sec:hubbard}. In \sref{sec:selfEnergy} and \sref{sec:chis_and_Us_model}, we recall the general functional derivative approach of Martin and Schwinger~\cite{Martin_1959, Kadanoff_1962} that allows us to find exact results and to set up the TPSC approach in the following section (\sref{sec:tpsc}).
\subsection{Hubbard model}
\label{sec:hubbard}
We study the one-band Hubbard model in dimension two or more
\begin{equation}
    H = \sum_{\mathbf{k},\sigma}\epsilon_{\mathbf{k}}c^{\dagger}_{\mathbf{k}\sigma}c_{\mathbf{k}\sigma} + U\sum_{i}n_{i\uparrow}n_{i\downarrow},
    \label{eq:hubbard}
\end{equation}
where $c^{(\dagger)}_{\mathbf{k}\sigma}$ annihilates (creates) an electron of spin $\sigma$ and wave vector $\mathbf{k}$, $n_{i\sigma}$ counts the number of electrons of spin $\sigma$ at site $i$, $U$ is the on-site repulsive interaction, and $\epsilon_{\mathbf{k}}$ is the bare band dispersion with wave vector $\mathbf{k}$. Working in units where $\hbar=k_B=1$, this dispersion is defined as
\begin{equation}
   \sum_{\mathbf{k},\sigma}\epsilon_{\mathbf{k}}c^{\dagger}_{\mathbf{k}\sigma}c_{\mathbf{k}\sigma}=\sum_{i,j,\sigma}t_{ij}c^{\dagger}_{i\sigma}c_{j\sigma},
   \label{eq:dispersion}
\end{equation}
where $t_{ij}$ is the hopping amplitude between sites $i$ and $j$. Throughout this paper, we focus on the $2D$ square lattice with first neighbour hopping $t$ only and set the lattice spacing to $a=1$, corresponding to the dispersion $\epsilon_\mathbf{k}=-2t(\cos(k_x)+\cos(k_y))$. Though most of the results shown are at half-filling ($n=1$), we also show some benchmarks away from half-filling in \sref{sec:results_doped}. We set $t=1$ as the unit of energy. The benchmarks in \sref{sec:results_doped} are provided in two dimensions. 

\subsection{Self-energy in the Hubbard model}
\label{sec:selfEnergy}
In this section, we derive an expression for the self-energy of the Hubbard model using the source field approach \cite{Martin_1959, Kadanoff_1962}. We start by defining a partition function $Z$ in the presence of a source field $\phi$
\begin{align}
    Z[\phi] = 
    \langle T_\tau e^{-c^{\dagger}_{\bar{\sigma}}(\bar{1})c_{\bar{\sigma}}(\bar{2})\phi_{\bar{\sigma}}(\bar{1},\bar{2})} \rangle,
 \label{eq:Z}
 \end{align}
where $T_\tau$ is the time-ordering operator and $\langle O\rangle = \mathrm{Tr}[Oe^{-\beta(H-\mu N)}]/\mathrm{Tr}[e^{-\beta(H-\mu N)}]$ is the thermodynamic average of the operator $O$ in the the grand-canonical ensemble. Moreover, we introduce the notation $(\mathbf{r}_1,\tau_1)\equiv (1)$. The bar denotes a sum over the spin, position and imaginary time. For instance we have, explicitly,
\begin{align}
    c_{\bar{\sigma}}(\bar{1})\phi(\bar{1},2)&=\\ \nonumber
    &\int_0^\beta d\tau_1 \sum_{\mathbf{r}_1}\sum_{\sigma_1} c_{\sigma_1}(\mathbf{r}_1,\tau_1)\phi_{\sigma_1,\sigma_2}(\mathbf{r}_1,\tau_1,\mathbf{r}_2,\tau_2). 
\end{align}
We use the partition function \eref{eq:Z} to define the Green's function in the presence of a source field
\begin{align}
    \mathcal{G}_\sigma(1,2)_\phi &= -\frac{\delta \mathrm{ln} Z[\phi]}{\delta \phi_{\sigma}(2,1)},
 \label{eq:G_phi}
 &= -\langle c_\sigma(1)c_\sigma^{\dagger}(2)\rangle_\phi,
 \end{align}
where the symbol $\delta$ denotes a functional derivative. Higher-order correlation functions are obtained from additional functional derivatives. Also, the thermodynamic average of an operator $O$ in the presence of the source field $\phi$ is defined as 
\begin{equation}
    \langle O \rangle_\phi = \frac{\langle T_\tau e^{-c^{\dagger}_{\bar{\sigma}}(\bar{1})c_{\bar{\sigma}}(\bar{2})\phi_{\bar{\sigma}}(\bar{1},\bar{2})} O\rangle }{Z[\phi]}.
\end{equation} 

The Green's function of \eref{eq:G_phi} is related to the usual Green's function by setting the source field $\phi$ to $0$. 

From the equations of motion for the Green's function in the presence of a source field, we obtain the Green's function through the Dyson equation and an exact expression (Schwinger-Dyson) for the self-energy~\cite{Kadanoff_1962}
\begin{align}
     \mathcal{G}^{-1}_\sigma(1,2)  &=  \mathcal{G}^{0^{-1}}_\sigma(1,2) - \phi(1,2) -\Sigma_\sigma(1,2)_\phi,
    \label{eq:Gphi_Dyson}
\end{align}
\begin{equation}
    \Sigma_\sigma(1,\bar{2})_\phi\mathcal{G}_\sigma(\bar{2},2)_\phi = U\langle T_{\tau}c^{\dagger}_\sigma(2)c^{\dagger}_{-\sigma}(1^+)c_{-\sigma}(1)c_{\sigma}(1)\rangle_\phi.
    \label{eq:SigmaPhi}
\end{equation}

We use the notation $(1^+) = (\mathbf{r}_1,\tau_1+0^+)$. 

\subsection{Spin and charge irreducible vertices}
\label{sec:chis_and_Us_model}

In the source field approach, the generalized susceptibilities are
\begin{align}
    \chi_{+}(1,3;2,4) &= \lim_{\phi\rightarrow 0} \sum_{\sigma,\sigma'}-\frac{\delta \mathcal{G}_\sigma(1,3)_\phi}{\delta \phi_{\sigma'}(4,2)}, \label{eq:chich_deriv}\\
    \chi_{-}(1,3;2,4) &= \lim_{\phi\rightarrow 0} \sum_{\sigma,\sigma'}-\sigma \sigma'\frac{\delta \mathcal{G}_\sigma(1,3)_\phi}{\delta \phi_{\sigma'}(4,2)}, \label{eq:chisp_deriv}
\end{align}
where the spin index $\sigma$ is equal to $\pm 1$ when used as a variable in the sum. The previous equations are obtained directly from the definition of $\mathcal{G}_\sigma(1,2)_\phi$ given in \eref{eq:G_phi}, which indeed leads to
\begin{align}
    \frac{\delta \mathcal{G}_\sigma(1,3)_\phi}{\delta \phi_{\sigma'}(4,2)} &= \mathcal{G}_\sigma(1,3)_\phi \mathcal{G}_{\sigma'}(2,4)_\phi \nonumber \\
    &- \langle T_\tau c_\sigma(1)c^\dagger_\sigma(3) c_{\sigma'}(2)c^\dagger_{\sigma'}(4) \rangle_\phi.
\end{align}
From spin rotational invariance, \eref{eq:chich_deriv} and \eref{eq:chisp_deriv} can be written as 
\begin{align}
    \chi_{+}(1,3;2,4) &= \lim_{\phi\rightarrow 0} -2\left [\frac{\delta \mathcal{G}_\uparrow(1,3)_\phi}{\delta \phi_{\uparrow}(4,2)} + \frac{\delta \mathcal{G}_\uparrow(1,3)_\phi}{\delta \phi_{\downarrow}(4,2)} \right ],\\
    \chi_{-}(1,3;2,4) &= \lim_{\phi\rightarrow 0} -2\left [\frac{\delta \mathcal{G}_\uparrow(1,3)_\phi}{\delta \phi_{\uparrow}(4,2)} - \frac{\delta \mathcal{G}_\uparrow(1,3)_\phi}{\delta \phi_{\downarrow}(4,2)} \right ].
    \label{eq:generalized_chis}
\end{align}
The relationship between the generalized susceptibilities defined in \eref{eq:chisp_deriv} and \eref{eq:chich_deriv} and the spin and charge susceptibilities is 
\begin{equation}
    \chi_{ch,sp}(1,2) = \chi_{+,-}(1,1^+;2,2^+).
\end{equation}

Expanding the equations for the generalized susceptibilities using spin rotational invariance and the definition of the Green's function in the presence of a source field \eref{eq:Gphi_Dyson}, we obtain
\begin{align}
    \chi_{ch}(1,2) &= -2\mathcal{G}_\sigma(1,2)\mathcal{G}_\sigma(2,1)\nonumber \\
    &+ \mathcal{G}_\sigma(1,\bar{3}) U_{ch}(\bar{3},\bar{4};\bar{5},\bar{6}) \chi_{ch}(\bar{5},\bar{6};2^+,2)\mathcal{G}_\sigma(\bar{4},1^+), \label{eq:chich_vertex}\\
    \chi_{sp}(1,2) &= -2\mathcal{G}_\sigma(1,2)\mathcal{G}_\sigma(2,1) \nonumber \\
    &- \mathcal{G}_\sigma(1,\bar{3}) U_{sp}(\bar{3},\bar{4};\bar{5},\bar{6})\chi_{sp}(\bar{5},\bar{6};2^+,2)\mathcal{G}_\sigma(\bar{4},1^+), \label{eq:chisp_vertex}
\end{align}
where the irreducible spin and charge vertices are defined as
\begin{align}
    U_{sp}(\bar{3},\bar{4};\bar{5},\bar{6}) &=  \frac{\delta \Sigma_\uparrow(\bar{3},\bar{4})}{\delta \mathcal{G}_{\downarrow}(\bar{5},\bar{6})} - \frac{\delta \Sigma_\uparrow(\bar{3},\bar{4})}{\delta \mathcal{G}_{\uparrow}(\bar{5},\bar{6})}, \label{eq:Usp_def}\\
    U_{ch}(\bar{3},\bar{4};\bar{5},\bar{6}) &= \frac{\delta \Sigma_\uparrow(\bar{3},\bar{4})}{\delta \mathcal{G}_{\uparrow}(\bar{5},\bar{6})} + \frac{\delta \Sigma_\uparrow(\bar{3},\bar{4})}{\delta \mathcal{G}_{\downarrow}(\bar{5},\bar{6})}. \label{eq:Uch_def}
\end{align}
Once we set the source field to zero, these expressions for $U_{sp}$ and $U_{ch}$ can, in general, be functions of three imaginary time differences (frequency) and of three position differences (wave vector). Assuming that they are local (in the next section), i.e. delta-functions in all time and position differences, we need to determine only two scalars, $U_{sp}$ and $U_{ch}$. 

Finally, we note that in the time and space invariant case, the spin and charge susceptibilities obey the exact local sum rules
\begin{align}
    \chi_{sp}(\mathbf{r}=0,\tau=0) &= \frac{T}{N}\sum_{\mathbf{q},iq_n} \chi_{sp}(\mathbf{q},iq_n), \\
    &= n - 2 \langle n_{\uparrow}n_{\downarrow} \rangle,
    \label{eq:def_sumrule_sp}
\end{align}

\begin{align}
\chi_{ch}(\mathbf{r}=0,\tau=0) &= \frac{T}{N}\sum_{\mathbf{q},iq_n} \chi_{ch}(\mathbf{q},iq_n), \\
    &= n + 2 \langle n_{\uparrow}n_{\downarrow} \rangle-n^2,
 \label{eq:def_sumrule_ch}
 \end{align}
 where we use the Fourier transforms with $q_n = 2n\pi T$  as bosonic Matsubara frequencies, with $\mathbf{q}$ as wave vectors in the Brillouin zone, with $N$ as the total number of sites in the system and $T$ the temperature. These expressions for the local spin and charge susceptibilities are obtained by enforcing the Pauli principle through $\langle n_\sigma^2 \rangle = \langle n_\sigma\rangle$.


\section{TPSC}
\label{sec:tpsc}

In this section, we recall some of the main properties of the TPSC approach. This method, which was first developed for the one-band Hubbard model, is valid in the weak to intermediate coupling regime. It is a conserving approach that respects both the Mermin-Wagner theorem and the Pauli principle \cite{Vilk_1997}. The starting point of the TPSC approach for the Hubbard model is the Schwinger-Dyson self-energy defined in \eref{eq:SigmaPhi}. We first impose that \eref{eq:SigmaPhi} is satisfied exactly at equal time and position, namely
\begin{equation}
    \Sigma_\sigma(1,\bar{2})_\phi\mathcal{G}_\sigma(\bar{2},1^+)_\phi = U\langle n_\uparrow(1) n_\downarrow(1) \rangle_\phi.
    \label{eq:SigmaExact}
\end{equation}
Next, we consider a Hartree-Fock like factorization of \eref{eq:SigmaPhi} when the point $2$ is different from the point $1$. We perform the factorization by  introducing a functional $A_\phi$
\begin{equation}
    \Sigma^{(1)}_\sigma(1,\bar{2})_\phi\mathcal{G}^{(1)}_\sigma(\bar{2},2)_\phi = A_\phi \mathcal{G}^{(1)}_{-\sigma}(1,1^+)_\phi \mathcal{G}^{(1)}_\sigma(1,2)_\phi.
    \label{eq:FactorAphi}   
\end{equation}
 The superscript $^{(1)}$ denotes the first level of approximation. The TPSC ansatz postulates that \eref{eq:SigmaExact} and \eref{eq:FactorAphi} must be satisfied simultaneously, in the spirit of Singwi~\cite{Singwi:1981} and Hedeyati-Vignale~\cite{Hedeyati:1989}. This means that the functional $A_\phi$ must be defined as 
\begin{equation}
    A_\phi = U \frac{\langle n_\uparrow(1) n_\downarrow(1) \rangle_\phi }{\langle n_\uparrow(1) \rangle_\phi\langle  n_\downarrow(1) \rangle_\phi}.
\end{equation}
We now compute the irreducible spin vertex defined in \eref{eq:Usp_def} using the first level of approximation for the self-energy. Setting the source field to zero after functional differentiation, we find
\begin{equation}
    U_{sp}(\bar{3},\bar{4};\bar{5},\bar{6}) =  A_{\phi=0} \delta(\bar{3}-\bar{5})\delta(\bar{3}-\bar{6})\delta(\bar{3}-\bar{4}).
\end{equation}
This leads to the following expression for the vertex $U_{sp}$, which is local in space and time
\begin{equation}
    U_{sp} = U\frac{\langle n_\uparrow n_\downarrow \rangle}{\langle n_\uparrow \rangle\langle n_\downarrow \rangle}.
    \label{eq:ansatz}
\end{equation}

Assuming that and $U_{ch}$ is also local, we obtain  RPA-like expressions for the spin and charge susceptibilities from \eref{eq:chich_vertex} and \eref{eq:chisp_vertex}
\begin{equation}
    \chi_{sp}(\mathbf{q},iq_n) = \frac{\chi^{(1)}(\mathbf{q},iq_n)}{1-\frac{U_{sp}}{2}\chi^{(1)}(\mathbf{q},iq_n)},
    \label{eq:chispqiqn}
\end{equation}

\begin{equation}
    \chi_{ch}(\mathbf{q},iq_n) = \frac{\chi^{(1)}(\mathbf{q},iq_n)}{1+\frac{U_{ch}}{2}\chi^{(1)}(\mathbf{q},iq_n)}.
    \label{eq:chichqiqn}
\end{equation}

The bubble $\chi^{(1)}$, evaluated at the first level of approximation, is 
\begin{equation}
    \chi^{(1)}(\mathbf{q},iq_n) = -2\frac{T}{N}\sum_{\mathbf{k},ik_n}\mathcal{G}^{(1)}_\sigma(\mathbf{k},ik_n)\mathcal{G}^{(1)}_\sigma(\mathbf{k}+\mathbf{q},ik_n+iq_n),
    \label{eq:chi1qiqn}
\end{equation}
where $k_n=(2n+1)\pi T$ are fermionic Matsubara frequencies and $\mathbf{k}$ are wave vectors in the first Brillouin zone.

The TPSC approach solves the Hubbard model through the self-consistency of the ansatz that leads to the definition of $U_{sp}$ (\eref{eq:ansatz}) and the sum rules for the spin and charges susceptibilities. Indeed, comparing the spin susceptibility sum rule \eref{eq:def_sumrule_sp} and the TPSC equation for the spin susceptibility \eref{eq:chispqiqn}, we find
\begin{align}
    \frac{T}{N}\sum_{\mathbf{q},iq_n} \frac{\chi^{(1)}(\mathbf{q},iq_n)}{1-\frac{U_{sp}}{2}\chi^{(1)}(\mathbf{q},iq_n)} &= n - 2\langle n_\uparrow n_\downarrow \rangle, \\
    &= n-2\frac{U_{sp}}{U}\langle n_\uparrow \rangle \langle n_\downarrow \rangle, \label{eq:selfcons}
\end{align}
where the second line comes from \eref{eq:ansatz}, which defines $U_{sp}$ from the double occupancy $\langle n_\uparrow n_\downarrow\rangle$. 

We first solve \eref{eq:selfcons} self-consistently for $U_{sp}$ and the double occupancy. Then, given the double occupancy, the sum rule on the charge susceptibility 
\begin{align}
    \frac{T}{N}\sum_{\mathbf{q},iq_n} \frac{\chi^{(1)}(\mathbf{q},iq_n)}{1+\frac{U_{ch}}{2}\chi^{(1)}(\mathbf{q},iq_n)} &= n + 2\langle n_\uparrow n_\downarrow \rangle - n^2. \label{eq:chich_selfcons}
\end{align}
gives us the value of $U_{ch}$. Since the expressions for the local spin and charge sum rules used within TPSC enforce the Pauli principle, the method itself respects it.
Since the self-energy in the first level of approximation $\Sigma_\sigma^{(1)}(1,2)=U_{sp}n_{-\sigma}\delta(\mathbf{r}_1-\mathbf{r}_2)\delta(\tau_1-\tau_2)$ is a constant, we use the noninteracting Lindhard function $\chi^{(0)}$ in the TPSC equations and absorb $\Sigma^{(1)}$ in the definition of the chemical potential. 

Within TPSC, in analogy with what is done for the electron gas, at the second level of approximation the self-energy is influenced by the spin and charge fluctuations. It enters the interacting Green's function through $\mathcal{G}^{(2)}=\left( \mathcal{G}^{(1)^{-1}} - \Sigma^{(2)} \right)^{-1}$ and is given by
\begin{align}
    \Sigma_\sigma^{(2)}(\mathbf{k},ik_n) &= Un_{-\sigma}\nonumber+\frac{T}{N}\frac{U}{8}\sum_{\mathbf{q},iq_n}\left [3U_{sp}\chi_{sp}(\mathbf{q},iq_n)\right.\nonumber \\
     &+\left. U_{ch}\chi_{ch}(\mathbf{q},iq_n)\right ] \mathcal{G}_\sigma^{(1)}(\mathbf{k}+\mathbf{q},ik_n+iq_n).
    \label{eq:selfEnergy2}
\end{align}
This expression for the self-energy \eref{eq:selfEnergy2} contains the contribution from the longitudinal \cite{Vilk_1997} and transverse \cite{Allen_2003} channels. 

This form satisfies exactly the Galitski-Migdal equation 
\begin{align}
    Tr\left[\Sigma^{(2)} \mathcal{G}^{(1)} \right] = U \langle n_{\uparrow} n_{\downarrow} \rangle,
    \label{eq:trsigmag_tpsc}
\end{align}
demonstrating consistency between one- and two-particle quantities. However, within TPSC, the equality in \eref{eq:trsigmag_tpsc} is not satisfied if one uses the interacting Green's function $\mathcal{G}^{(2)}$ instead of the non-interacting one $\mathcal{G}^{(1)}$. The deviation between the trace of $\Sigma^{(2)} \mathcal{G}^{(i)}$ with the noninteracting ($i=1$) and the interacting ($i=2$) Green's functions can be used as an internal consistency check of the approach~\cite{Vilk_1997}.


\section{TPSC+}
\label{sec:tpsc+}

The main aim of the TPSC+ approach is to improve the results deep in the renormalized classical regime where, as will be shown in \sref{sec:crr_tpscplus}, the TPSC approach fails. We start by introducing the formulation of two variants of the TPSC+ approach in \sref{sec:intro_tpscplus}, called TPSC+ and TPSC+SFM. Then, we show that the methods respect the Mermin-Wagner theorem in \sref{sec:mw_tpscplus}, that they are valid in the renormalized classical regime in \sref{sec:crr_tpscplus}, and that they recover a generalized Stoner criterion with a renormalized interaction in the antiferromagnetic phase in \sref{sec:stoner_tpscplus}. Moreover, we show  that the TPSC+ approach is consistent with respect to one- and two-particle quantities in \sref{sec:sigmag_tpscplus}. In \sref{sec:pg_tpscplus}, we show that the methods predict an antiferromagnetic pseudogap in the $2D$ Hubbard model. Finally, we comment on the limitations of the TPSC+ approach in \sref{sec:lim_tpscplus}, where we show that it violates spin and charge conservation laws as well as the f-sum rule, whereas its variant TPSC+SFM does not. 

\subsection{Formulation of the approach}
\label{sec:intro_tpscplus}
The two TPSC+ approaches introduced here are based on the same considerations we introduced for the TPSC approach. In the TPSC+ approach, the self-energy $\Sigma^{(2)}$ that enters $\mathcal{G}^{(2)}$ is defined as in \eref{eq:selfEnergy2}. However, the spin and charge susceptibilities are no longer defined with the non-interacting susceptibility $\chi^{(1)}$, but instead as
\begin{align}
    \chi_{sp}(\mathbf{q},iq_n) = \frac{\chi^{(2)}(\mathbf{q},iq_n)}{1-\frac{U_{sp}}{2}\chi^{(2)}(\mathbf{q},iq_n)},
    \label{eq:chisp2}\\
    \chi_{ch}(\mathbf{q},iq_n) = \frac{\chi^{(2)}(\mathbf{q},iq_n)}{1+\frac{U_{ch}}{2}\chi^{(2)}(\mathbf{q},iq_n)}.
    \label{eq:chich2}
\end{align}
The spin and charge irreducible vertices are computed in the same way as in the TPSC approach, namely through the self-consistency with the local sum rules and the TPSC ansatz $U_{sp}=U\langle n_\uparrow n_\downarrow \rangle / \langle n_\uparrow \rangle \langle n_\downarrow \rangle$. The distinction between TPSC, TPSC+ and TPSC+SFM comes from the asymmetric form of the partially-dressed susceptibility $\chi^{(2)}$ that we consider here. In TPSC+, it is defined as
\begin{align}
    \chi^{(2)}_{\mathrm{TPSC+}}(\mathbf{q},iq_n) = &-\frac{T}{N}\sum_{\mathbf{k},ik_n}\left (\mathcal{G}_{\sigma}^{(2)}(\mathbf{k},ik_n)\mathcal{G}_{\sigma}^{(1)}(\mathbf{k}+\mathbf{q},ik_n+iq_n)\right.\nonumber \\
     &+\left. \mathcal{G}_{\sigma}^{(2)}(\mathbf{k},ik_n)\mathcal{G}_{\sigma}^{(1)}(\mathbf{k}-\mathbf{q},ik_n-iq_n) \right ).
    \label{eq:chi2}
\end{align}
In TPSC+SFM, the partially-dressed susceptibility takes a different form
\begin{widetext}
\begin{align}
    \chi^{(2)}_{\mathrm{TPSC+SFM}}(\mathbf{q},iq_n) &= \left \{ \begin{matrix} -\frac{T}{N}\sum_{\mathbf{k},ik_n}\left (\tilde{\mathcal{G}}_{\sigma}^{(2)}(\mathbf{k}-\mathbf{q},ik_n)\mathcal{G}_{\sigma}^{(1)}(\mathbf{k},ik_n) + \tilde{\mathcal{G}}_{\sigma}^{(2)}(\mathbf{k}+\mathbf{q},ik_n)\mathcal{G}_{\sigma}^{(1)}(\mathbf{k},ik_n) \right ), & q_n=0\\
     \chi^{(1)}(\mathbf{q},iq_n) & q_n\neq 0.
    \end{matrix} \right .
    \label{eq:chi2m}
\end{align}
\end{widetext}
In \eref{eq:chi2} and \eref{eq:chi2m}, the Green's function $\mathcal{G}^{(1)}$ is the non-interacting Green's function, and the Green's function $\mathcal{G}^{(2)}$ is the interacting Green's function that includes the complete TPSC self-energy defined previously in \eref{eq:selfEnergy2}. The Green's function $\tilde{\mathcal{G}}^{(2)}$ is an interacting Green's function in which the self-energy only contains the contribution from the longitudinal spin fluctuations. Hence, the distinction between the interacting Green's functions $\mathcal{G}^{(2)}$ and $\tilde{\mathcal{G}}^{(2)}$ is
\begin{widetext}
\begin{align}
    \mathcal{G}^{(2)} \Rightarrow \Sigma^{(2)}(\mathbf{k},ik_n) &= \frac{T}{N}\frac{U}{8}\sum_{\mathbf{q},iq_n}\left [3U_{sp}\chi_{sp}(\mathbf{q},iq_n)+ U_{ch}\chi_{ch}(\mathbf{q},iq_n)\right ] \mathcal{G}_\sigma^{(1)}(\mathbf{k}+\mathbf{q},ik_n+iq_n). \\
    \tilde{\mathcal{G}}^{(2)} \Rightarrow \tilde{\Sigma}^{(2)}(\mathbf{k},ik_n) &= \frac{T}{N}\frac{U}{4}\sum_{\mathbf{q},iq_n} U_{sp}\chi_{sp}(\mathbf{q},iq_n) \mathcal{G}_\sigma^{(1)}(\mathbf{k}+\mathbf{q},ik_n+iq_n). \label{eq:selfEnergySFM}
\end{align}
\end{widetext}
In the Green's functions $\mathcal{G}^{(2)}$ and $\tilde{\mathcal{G}}^{(2)}$, the chemical potential is chosen so that the total density $n$ is kept constant. This means that the partially-dressed susceptibilities $\chi^{(2)}$ calculated from TPSC+ and TPSC+SFM obey the same sum rule as the noninteracting correlation function $\chi^{(1)}$
\begin{align}
    \chi^{(1)}(\mathbf{r}=0,\tau=0) &= \chi^{(2)}(\mathbf{r}=0,\tau=0),\nonumber\\
    &= n - \frac{n^2}{2}.
\end{align}

We note that the final Green's function obtained in the TPSC+SFM approach is still the one that includes both spin and charge fluctuations, as defined in \eref{eq:selfEnergy2}. Only the self-energy used in the calculation of the partially-dressed susceptibility takes the form defined in \eref{eq:selfEnergySFM}. We also remark that the discontinuity in the partially-dressed susceptibility introduced in \eref{eq:chi2m} could be an issue for the analytic continuation to real frequencies.

Both TPSC+ approaches are self-consistent in two ways: (a) The self-consistency between $U_{sp}$ and the double occupancy through the sum rule and the TPSC ansatz is still present in the extended approaches, and (b) The self-energy $\Sigma_\sigma^{(2)}$, the Green's function $\mathcal{G}^{(2)}_\sigma$ and the partially-dressed susceptibility $\chi^{(2)}$ all depend self-consistently on each other and can be calculated through an iterative process.

This approach is analogous to the pairing approximation ($GG_0$ theory) for the pair susceptibility introduced by Kadanoff and Martin \cite{Kadanoff_Martin_1961, Chen_2005, Boyack_2018}. Though we will rigorously justify the approach in the following sections, we now provide a phenomenological justification for the use of a partially-dressed susceptibility $\chi^{(2)}$, which was also introduced in Appendix D.7 of Ref. \cite{Schafer_2021}. As detailed in Section \ref{sec:chis_and_Us_model}, in the source field approach the susceptibilities are obtained from functional derivatives of the Green's function. Using the identity $G(1,\bar{3})G^{-1}(\bar{3},2)=\delta(1-2)$, these susceptibilities can be written as
\begin{equation}
    \frac{\delta G}{\delta \phi} = - G\frac{\delta G^{-1}}{\delta \phi} G,
    \label{eq:deltaGdeltaPhi}
\end{equation}
where $\frac{\delta G^{-1}}{\delta \phi}$ is the vertex. Assuming a quasiparticle picture, the Green's function can be expressed as a function of the quasiparticle weight $Z$ and the non-interacting Green's function $G_0$, namely $G = ZG_0$. Hence, in this approximation, the functional derivative of the Green's function becomes analogous to susceptibilities introduced in Eqs. \ref{eq:chisp2}, \ref{eq:chich2} and \ref{eq:chi2},
\begin{equation}
    \frac{\delta G}{\delta \phi} = - G_0\frac{\delta G_0^{-1}}{\delta \phi} G.
\end{equation}
This is reminiscent of the cancellation between quasiparticle renormalization in the Green function and in the vertex that occurs in Landau Fermi liquid theory. 

\subsection{Mermin-Wagner theorem}
\label{sec:mw_tpscplus}
Here, we show that the TPSC+ approach and the TPSC+SFM variant respect the Mermin-Wagner theorem and more specifically that it prevents antiferromagnetic phase transitions at finite temperatures in two dimensions. The proof is the same as in the case of the TPSC approach detailed in Ref. \cite{Vilk_1997}. We consider a regime where the spin correlation length is large. In this regime, the spin susceptibility can be expanded in an Ornstein-Zernicke form at zero frequency and for wave vectors near the antiferromagnetic wave vector $\mathbf{Q}$
\begin{equation}
    \chi_{sp}(\mathbf{q}\simeq\mathbf{Q},0) \simeq \frac{2}{U_{sp}\xi_0^2} \frac{1}{q^2 + \xi_{sp}^{-2}},
    \label{eq:chispOZ}
\end{equation}
where the bare particle-hole correlation length is defined as
\begin{equation}
    \xi_0^2 = \frac{-1}{2\chi^{(2)}(\mathbf{Q}, 0)}\left .\frac{\partial^2\chi^{(2)}(\mathbf{q},0)}{\partial q_x^2}\right|_{\mathbf{q}=\mathbf{Q}},
\end{equation}
and the spin correlation length is
\begin{equation}
    \xi_{sp} = \xi_0\sqrt{\frac{U_{sp}}{\frac{2}{\chi^{(2)}(\mathbf{Q},0)}-U_{sp}}}.
\end{equation}
The sum rule for the spin susceptibility can be rewritten as
\begin{equation}
    n - 2\langle n_\uparrow n_\downarrow \rangle - C = \frac{2}{U_{sp}\xi_0^2} \sum_{\mathbf{q}} \frac{1}{q^2 + \xi_{sp}^{-2}},
    \label{eq:verifmw}
\end{equation}
where the constant $C$ contains all the non-zero Matsubara frequency contributions to the sum rule. Hence, the left handside of the previous equation is finite. We now focus on the right handside and transform the sum to an integral
\begin{equation}
    \sum_{\mathbf{q}}\chi_{sp}(\mathbf{q}\simeq\mathbf{Q},0) \simeq \frac{2}{U_{sp}\xi_0^2} \int \frac{d^dq}{(2\pi)^d} \frac{1}{q^2 + \xi_{sp}^{-2}},
\end{equation}
where $d$ is the spatial dimension. A more detailed analysis with power-law corrections can be found in Ref.~\cite{Roy:2008}.

We first consider the $d=3$ case, where $d^3q = q^2\sin\theta dq d\theta d\phi$. Assuming that an antiferromagnetic order is possible, we take the spin correlation length $\xi_{sp}$ to infinity. We obtain
\begin{equation}
   \int d^3q \frac{1}{q^2 + \xi_{sp}^{-2}} \propto \int_0^\Lambda dq,
\end{equation}
where $\Lambda$ is a finite cutoff to take into account the regime of validity of the Ornstein-Zernicke form of the spin susceptibility. Hence, in $d=3$, both sides of \eref{eq:verifmw} take finite values even in the limit where $\xi_{sp}\rightarrow \infty$, which means that an antiferromagnetic phase transition is possible at finite temperature. The same is true for $d>3$. 

The $d=2$ case, for which $d^2q = qdqd\phi$, is different. Indeed, taking $\xi_{sp}\rightarrow \infty$ in $d=2$ yields
\begin{equation}
    \int d^2q \frac{1}{q^2 + \xi_{sp}^{-2}} \propto \int_0^\Lambda \frac{dq}{q},
 \end{equation}
which diverges logarithmically at $q\rightarrow 0$ and leads to a contradiction in \eref{eq:verifmw} because its left hand-side remains finite. Hence, in $d=2$, the spin correlation length obtained from the TPSC+ approach never reaches an infinite value at finite temperature, in agreement with the Mermin-Wagner theorem.

\subsection{Validity in the renormalized classical regime}
\label{sec:crr_tpscplus}
The onset of the renormalized-classical regime is signaled by the characteristic spin fluctuation frequency $\omega_{SF} \propto \xi_{sp}^{-2}$ becoming smaller than the temperature $T$ \cite{chakravarty_two-dimensional_1989,Vilk_1997}. In this regime the spin correlation length grows exponentially. At lower temperature, the Vilk criterion becomes satisfied. The various crossovers associated with these phenomena have been thoroughly discussed in Ref.~\onlinecite{Schafer_2021}. One of the main limitations of the TPSC approach reviewed in \sref{sec:tpsc} is that it is not valid deep in the renormalized classical regime of the $2D$ Hubbard model.  In this section, we discuss how the TPSC+ methods are valid in this regime whereas TPSC is not.

We consider the half-filled case of the $2D$ Hubbard model with nearest-neighbor hopping only. 
As shown in \sref{sec:mw_tpscplus}, the TPSC and the TPSC+ approaches satisfy the Mermin-Wagner theorem, constraining the spin susceptibility at $\mathbf{Q}=(\pi,\pi)$ to finite values at finite temperature in $2D$. With the RPA-like spin susceptibilities of TPSC and TPSC+, this imposes a condition on the value of $U_{sp}$
\begin{equation}
    U_{sp} < \frac{2}{\chi^{(i)}(\mathbf{Q},0)},
    \label{eq:UspCondition}
\end{equation}
where $i=1$ ($i=2$) in the case of the TPSC (TPSC+) approach. 

\begin{figure}
    \centering
    \includegraphics[width=\columnwidth]{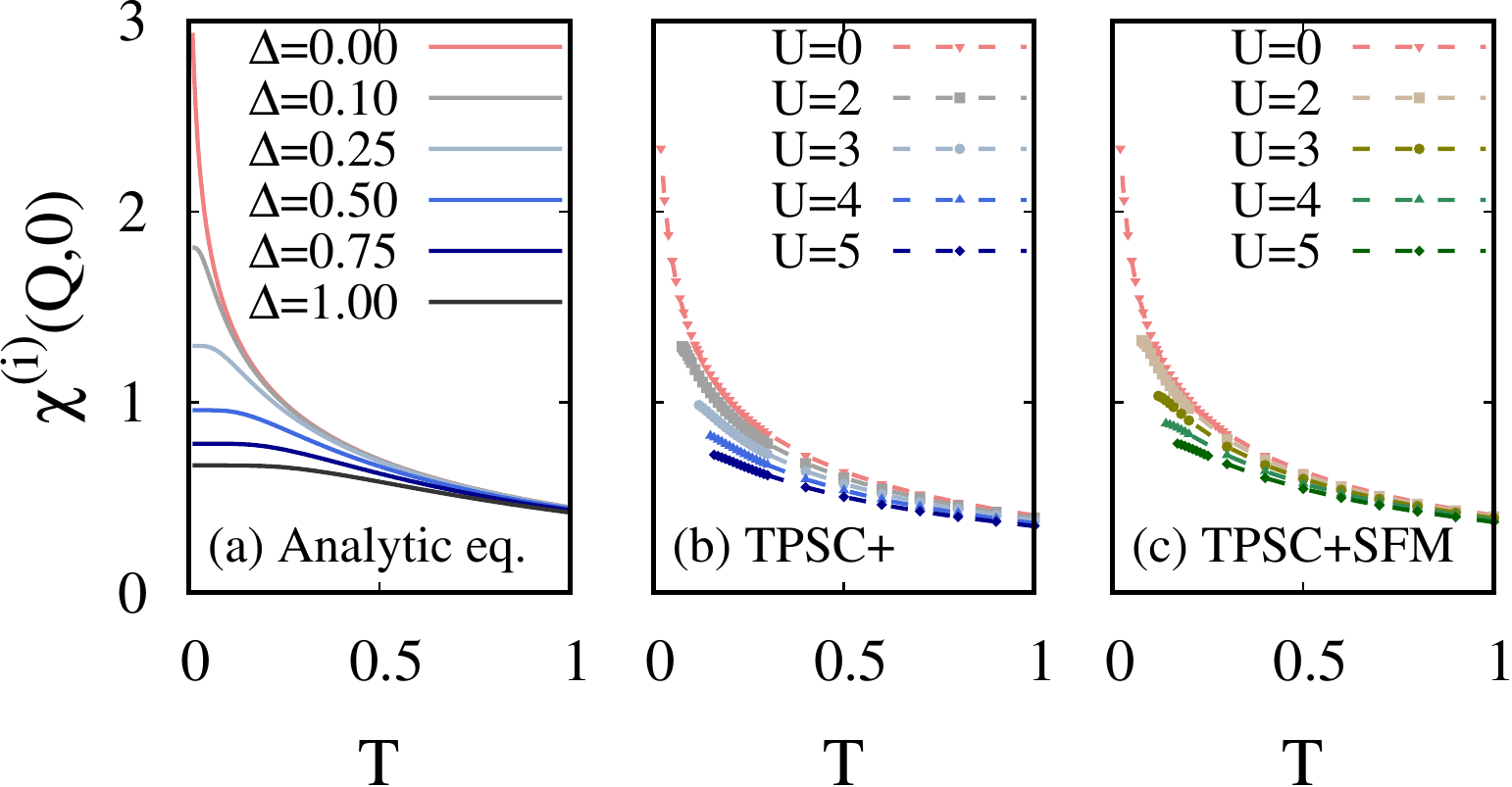} 
    \caption{Lindhard functions $\chi^{(1)}(\mathbf{Q},0)$ (in red) and $\chi^{(2)}(\mathbf{Q},0)$ (in blue and green) as a function of the temperature $T$. Lines in the panel (a) are obtained from the numerical integration of \eref{eq:chi0Q} for $i=1,~\Delta=0$ and of \eref{eq:chi2Q} for $i=2,~\Delta>0$. Dots in the panels (b) and (c) are obtained from the numerical calculation of (b) \eref{eq:chi2}  for TPSC+  and  (c) \eref{eq:chi2m} for TPSC+SFM. The dashed lines are an interpolation. The calculations are done for the $2D$ square lattice, at half-filling, with nearest-neighbour hopping only. }
    \label{fig:chi0_vs_chi2}
\end{figure}

We first consider the TPSC case. In the specific example mentioned earlier, the zero-frequency Lindhard function at $\mathbf{Q}=(\pi,\pi)$ is 
\begin{equation}
    \chi^{(1)}(\mathbf{Q},0) = \int d\epsilon \rho(\epsilon)\frac{\mathrm{tanh}(\epsilon/2T)}{\epsilon},
    \label{eq:chi0Q}
\end{equation}
where the density of states $\rho(\epsilon)$ is
\begin{equation}
    \rho(\epsilon) =\frac{1}{2\pi^2 t}K\left [ \sqrt {1-\left (\frac{\epsilon}{4t} \right)^2}\right] ,
    \label{eq:dos}
\end{equation}
with $K$ the complete elliptic integral of the first kind \cite{Benard_1993}.
The integral \eref{eq:chi0Q} can be solved numerically as a function of the temperature $T$. In \fref{fig:chi0_vs_chi2}, we show that there is a divergence in $\chi^{(1)}(\mathbf{Q},0)$ as $T$ goes to $0$ both from the numerical integration of \eref{eq:chi0Q} (panel (a), red curve) and from the numerical evaluation of \eref{eq:chi1qiqn} (panels (b) and (c), red dots). This divergence can also be shown from an analytic evaluation of \eref{eq:chi0Q}. Indeed, in the specific case we study, there is a van Hove singularity in the density of states at the Fermi level ($\epsilon_F=0$). More specifically, the dominant contribution to $K$ is
\begin{equation}
    K\left [ \sqrt {1-x^2}\right] \rightarrow -\ln{x},~~~x\rightarrow 0.
\end{equation}
The integral \eref{eq:chi0Q} can then be written, with $u=\epsilon/2T$,
\begin{align}
    \chi^{(1)}(\mathbf{Q},0) &\sim \left [ -\int_1^{\Lambda/2T} du \frac{\ln{u}}{u} + C \right ],\nonumber \\
    &\sim - \ln^2\frac{\Lambda}{2T},
\end{align}
where $C$ includes the less singular terms that contribute to the integral, and $\Lambda$ is a high energy cutoff. Hence, the Lindhard function diverges as the square of a logarithm at $T\rightarrow0$ for this model.

\begin{figure}
    \centering
    \includegraphics[width=\columnwidth]{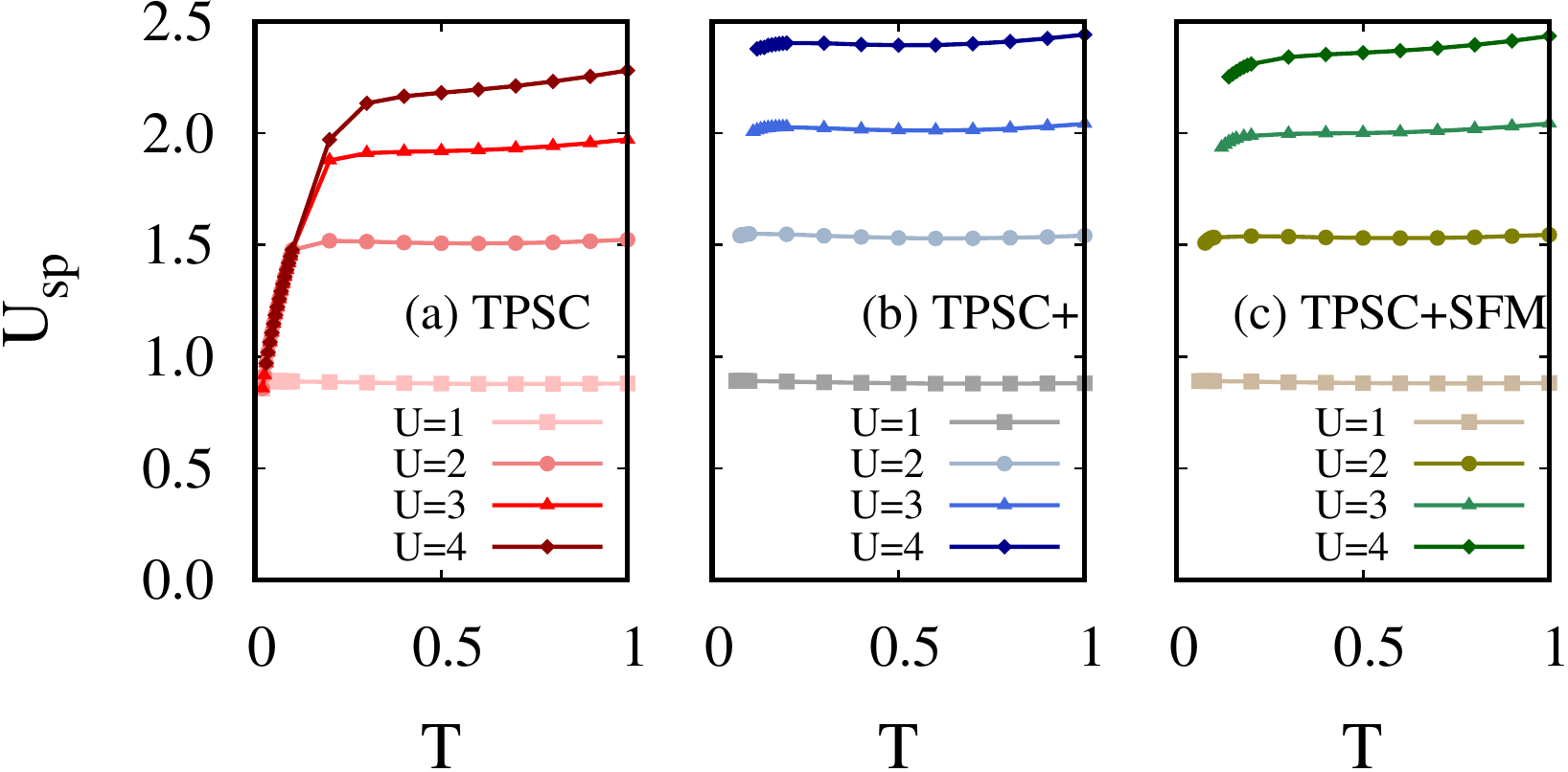} 
    \caption{Irreducible spin vertex $U_{sp}$ obtained from (a) the TPSC, (b) TPSC+ and (c) the TPSC+SFM calculations, for $U=1,~2,~3$ and $4$. The TPSC approach predicts an unphysical drop of $U_{sp}$ as the temperature goes to zero. In contrast, the values of $U_{sp}$ obtained with TPSC+ and TPSC+SFM show no significant decrease at low temperatures in the domain of convergence of the methods. The calculations are done for the $2D$ square lattice, at half-filling, with nearest-neighbour hopping only.}
    \label{fig:Usp_vs_T}
\end{figure}

From the bound on $U_{sp}$ emerging due to the Mermin-Wagner theorem \eref{eq:UspCondition}, we conclude that, in the TPSC approach, the spin vertex $U_{sp}$ must go to $0$ as $T$ goes to $0$ in order to respect the Mermin-Wagner theorem. In panel (a) of \fref{fig:Usp_vs_T}, we show $U_{sp}$ as a function of the temperature for $U=1,~2,~3$ and $4$ obtained from the TPSC calculation for the $2D$ Hubbard model with nearest-neighbor hopping only. All cases show a drop in the value of $U_{sp}$ as the temperature goes to zero. This drop signals the entry in the renormalized classical regime. Moreover, since the value of $U_{sp}$ is then limited by the value of the non-interacting Lindhard function, the spin vertex in this regime becomes independent of $U$.

We now show that the TPSC+ approach does not encounter this issue. We first need to evaluate $\chi^{(2)}(\mathbf{Q},0)$ in the renormalized classical regime where spin fluctuations are strong, starting with the computation of the self-energy that enters the interacting Green's function $\mathcal{G}^{(2)}$. We only consider the contributions from spin fluctuations and use the Ornstein-Zernicke form of the spin susceptibility. This approximation is valid for both the TPSC+ and the TPSC+SFM methods. In 2 dimensions, we find
\begin{equation}
    \Sigma(\mathbf{k},ik_n) \simeq \frac{3UT}{4\xi_0^2} \int \frac{d^{2}q}{(2\pi)^2} \frac{1}{q^2 + \xi_{sp}^{-2}} \frac{1}{ik_n-\epsilon_{\mathbf{k}+\mathbf{Q}+\mathbf{q}}+\mu^{(1)}}.
    \label{eq:selfOZ}
\end{equation}
Since the spin correlation length $\xi_{sp}$ is very large in this regime, the term $1/(q^2+\xi_{sp}^{-2})$ is non-negligible only in the limit where $q$ is small. We also note that, for values of the wave vector that lie outside of the Fermi surface, the term $\epsilon_{\mathbf{k}+\mathbf{Q}+\mathbf{q}}+\mu^{(1)}$ is non-zero. To compute $\chi^{(2)}$, we must perform a sum over the whole Brillouin zone. For all these reasons, we can neglect the $\mathbf{q}$ dependence of $\epsilon_{\mathbf{k}+\mathbf{Q}+\mathbf{q}}$ in the above expression. This leads to
\begin{equation}
    \Sigma(\mathbf{k},ik_n) = \frac{\Delta^2}{ik_n-\epsilon_{\mathbf{k}+\mathbf{Q}}+\mu^{(1)}},
\end{equation}
where $\Delta^2$ is defined as the temperature dependent quantity
\begin{equation}
\label{eq:Delta}
    \Delta^2 = \frac{3UT}{4\xi_0^2} \int \frac{d^{2}q}{(2\pi)^2} \frac{1}{q^2 + \xi_{sp}^{-2}}.
\end{equation}
We now obtain an expression for $\chi^{(2)}$ in the limit $T\rightarrow 0 $ in the renormalized classical regime
\begin{align}
    \chi^{(2)}(\mathbf{Q},0) &= -\frac{T}{N}\sum_{\mathbf{k},ik_n} \left [ \mathcal{G}^{(2)}(\mathbf{k},ik_n)\mathcal{G}^{(1)}(\mathbf{k}+\mathbf{Q},ik_n)\right. \nonumber \\
     &+ \left .\mathcal{G}^{(2)}(\mathbf{k},ik_n)\mathcal{G}^{(1)}(\mathbf{k}-\mathbf{Q},ik_n)\right],\\
     &= -\frac{T}{N}\sum_{\mathbf{k},ik_n} \frac{2}{(ik_n-\varepsilon^{(1)}_{\mathbf{k}+\mathbf{Q}})(ik_n-\varepsilon_{\mathbf{k}}^{(2)})-\Delta^2},
\end{align}
where we used the equality $\epsilon_{\mathbf{q}+\mathbf{Q}}=\epsilon_{\mathbf{q}-\mathbf{Q}}$, and defined $\varepsilon^{(i)}_{\mathbf{k}}=\epsilon_{\mathbf{k}}-\mu^{(i)}$. The sum over discrete Matsubara frequencies can be performed analytically. We find
\begin{equation}
    \chi^{(2)}(\mathbf{Q},0) = -\frac{2}{N}\sum_{\mathbf{k}} \frac{f(E_{\mathbf{k}}^+) - f(E_{\mathbf{k}}^-) }{E_{\mathbf{k}}^+-E_{\mathbf{k}}^-},
    \label{eq:chi2_diffE}
\end{equation}
where $f(\epsilon)$ is the Fermi-Dirac distribution. The energies $E_{\mathbf{k}}^{\pm}$ are defined as 
\begin{equation}
    E_{\mathbf{k}}^{\pm} = \frac{1}{2}\left ( \varepsilon^{(2)}_{\mathbf{k}} + \varepsilon^{(1)}_{\mathbf{k}+\mathbf{Q}} \pm \sqrt{\left( \varepsilon^{(2)}_{\mathbf{k}}  - \varepsilon ^{(1)}_{\mathbf{k}+\mathbf{Q}}\right)^2+4\Delta^2} \right ).
    \label{eq:eplusmoins}
\end{equation}
So far, this is a general result that can be applied to cases outside of perfect nesting. We transform the sum over $\mathbf{k}$ into an integral and obtain, for the case of perfect nesting where $\mu^{(1)}=\mu^{(2)}=0$,
\begin{equation}
    \chi^{(2)}(\mathbf{Q},0) = \int d\epsilon \rho(\epsilon) \frac{\tanh(\sqrt{\epsilon^2+\Delta^2}/2T) }{\sqrt{\epsilon^2+\Delta^2}},
    \label{eq:chi2Q}
\end{equation}
which is the analogue of \eref{eq:chi0Q}. Like for $\chi^{(1)}(\mathbf{Q},0)$, we solve \eref{eq:chi2Q} for $\chi^{(2)}(\mathbf{Q},0)$ through a numerical integration and vary the value of $\Delta$. Panel (a) of \fref{fig:chi0_vs_chi2} compares our results for $\chi^{(1)}(\mathbf{Q},0)$ and $\chi^{(2)}(\mathbf{Q},0)$ from numerical integration. The presence of the self-energy through $\Delta^2$ suppresses the divergence at low temperature progressively as we increase $\Delta$. As $T$ decreases, $\chi^{(2)}(\mathbf{Q},0)$ saturates at a finite value instead of diverging as $\chi^{(1)}$ does. Consequently, the criterion \eref{eq:UspCondition} can be satisfied with a finite, non-zero value of $U_{sp}$ in the TPSC+ approach since the correlation function $\chi^{(2)}$ remains finite.

In panel (b) of \fref{fig:chi0_vs_chi2}, we show our results for $\chi^{(1)}(\mathbf{Q},0)$, and also for $\chi^{(2)}(\mathbf{Q},0)$ obtained with TPSC+ calculations. Panel (c) of the same figure shows the results obtained with TPSC+SFM calculations. Though we were not able to obtain converged results at very low temperatures, we see that both the TPSC+ and the TPSC+SFM forms of $\chi^{(2)}(\mathbf{Q},0)$ are suppressed increasingly as $U$ increases. This is consistent with the behavior seen in panel (a) of \fref{fig:chi0_vs_chi2} from the analytic, approximate expression of $\chi^{(2)}$.

We show our calculated values of $U_{sp}$ with TPSC+ in panel (b) of \fref{fig:Usp_vs_T}, as a function of the temperature for $U=1,~2,~3$ and $4$ for the $2D$ Hubbard model with nearest-neighbor hopping only. Panel (c) of the same figure shows the results from TPSC+SFM calculations. In contrast to the TPSC case, no sharp decrease of the value of $U_{sp}$ obtained with TPSC+ and TPSC+SFM can be seen at low temperatures. $U_{sp}$ remains strongly $U$-dependent. The slight downturn observed for all values of $U$ at low temperatures can be understood through the proportionality between the spin vertex and the double occupancy due to the TPSC ansatz. In the weak correlation regime of the $2D$ Hubbard model, a decrease in the double occupancy is observed as the temperature goes down towards zero due to an increase in antiferromagnetic correlations which in turn lead to an increase of the local moment and a corresponding decrease in double occupancy~\cite{Vilk:1996, Vilk_1997, Paiva_2001, Schafer_2021}. 

\begin{figure}
    \centering
    \includegraphics[width=\columnwidth]{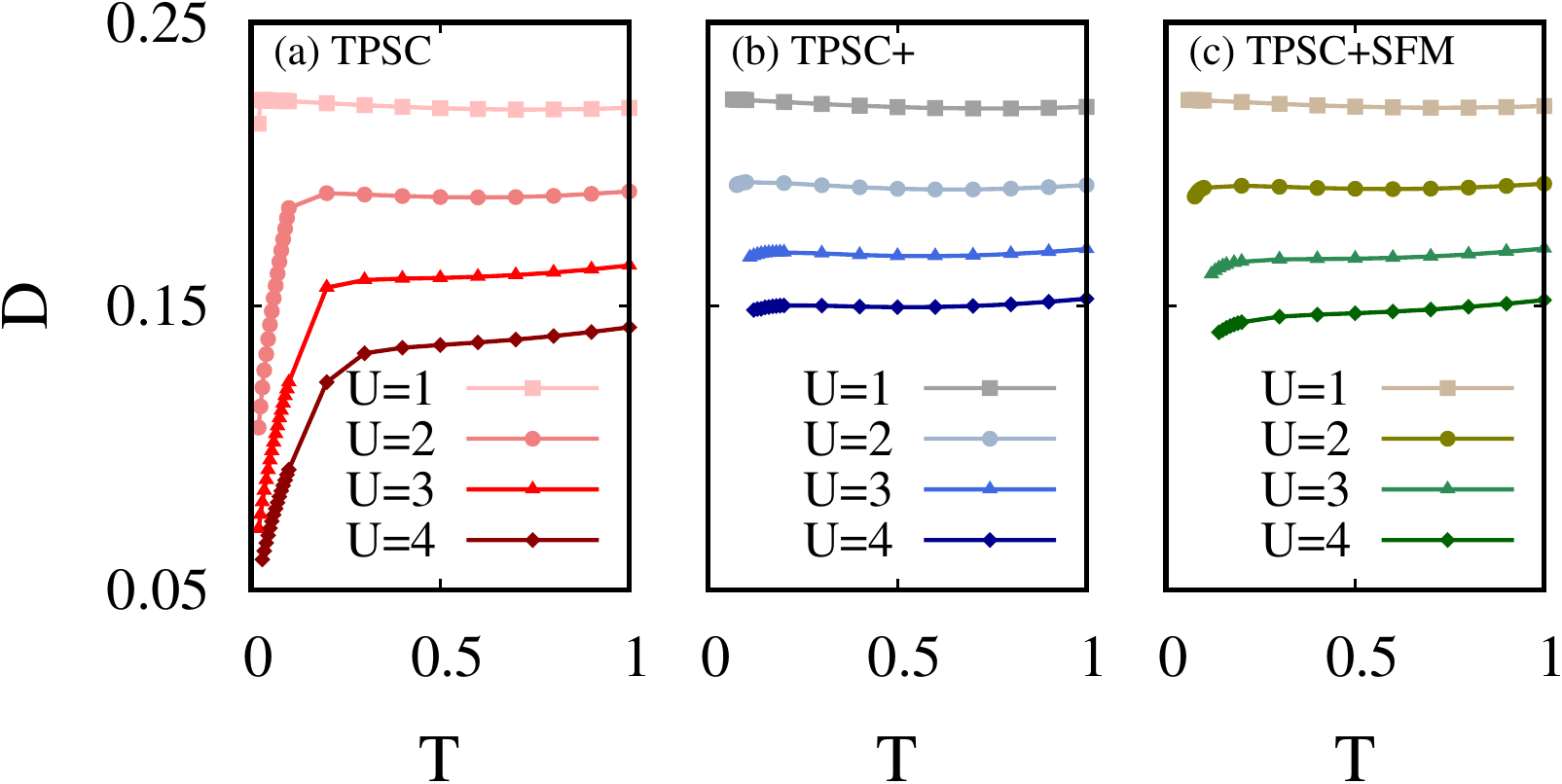} 
    \caption{Double occupancy $D=\langle n_\uparrow n_\downarrow\rangle$ obtained (a) from the TPSC, (b) TPSC+ and (c) TPSC+SFM (c) calculations, for $U=1,~2,~3$ and $4$. The TPSC approach predicts an unphysical drop of $D$ as the temperature goes to zero. In contrast, the double occupancy computed with TPSC+ and TPSC+SFM does not decrease significantly with the temperature in the domain of convergence of the methods. The calculations are done for the $2D$ square lattice, at half-filling, with nearest-neighbour hopping only.}
    \label{fig:docc_vs_T}
\end{figure}

A comparison of the double occupancies obtained with the TPSC, TPSC+ and TPSC+SFM approaches is shown in \fref{fig:docc_vs_T}. In panel (a), the double occupancy calculated with TPSC drops sharply towards zero for all values of $U$ considered. The drop occurs at higher temperature when $U$ increases since the entry in the renormalized classical regime has the same behavior. Comparison ~\cite{Kyung:2003a} with DCA calculations~\cite{Moukouri:2001} confirms this. However, the drop towards zero is unphysical.   
In contrast, the double occupancy computed with TPSC+ and TPSC+SFM shown in panels (b) and (c) exhibits the expected physical decrease as the temperature decreases, and in a less pronounced way than the TPSC case.

\subsection{Renormalized Stoner criterion}
\label{sec:stoner_tpscplus}
One of the main advantages of the TPSC+ methods is that in {\it two dimensions} they give results for the paramagnetic pseudogap phase all the way to, and including, the zero-temperature long-range antiferromagnetic phase. 

In this section then, we focus on the generalized Stoner criterion that leads to the AFM phase transition in the TPSC approach. At the Néel temperature, which is $T_N=0$ in $d=2$ but might be finite in higher dimensions, the spin susceptibility diverges according to the criterion
\begin{equation}
    \frac{2}{U_{sp}} = \chi^{(2)}(\mathbf{Q},0).
    \label{eq:chi2Usp}
\end{equation}
Substituting the criterion  \eref{eq:chi2Usp} in the expression  we obtained for $\chi^{(2)}(\mathbf{Q},0)$ in the previous section, \eref{eq:chi2_diffE}, we find the generalized Stoner criterion obtained from the TPSC+ and TPSC+SFM approaches at $T_N$ 
\begin{equation}
    \frac{1}{U_{sp}} = -\frac{1}{N}\sum_{\mathbf{k}} \frac{f(E_{\mathbf{k}}^+) - f(E_{\mathbf{k}}^-) }{E_{\mathbf{k}}^+-E_{\mathbf{k}}^-}
\end{equation}
with $E_{\pm}$ given by~\eref{eq:eplusmoins}.

In the specific case of two dimensions, where $T_N=0$, this becomes
\begin{equation}
    \frac{1}{U_{sp}} = -\frac{1}{N}\sum_{\mathbf{k}} \frac{\theta(-E_{\mathbf{k}}^+) - \theta(-E_{\mathbf{k}}^-) }{E_{\mathbf{k}}^+-E_{\mathbf{k}}^-},
\end{equation}
where $\theta(x)$ is the Heaviside function. Both results are analogous to the mean-field Hartree-Fock gap equation in the antiferromagnetic state \cite{Kusko_2002}, but with the renormalized spin vertex $U_{sp}$ instead of the bare $U$. From the definition of the energies $E^{\pm}_\mathbf{k}$ in \eref{eq:eplusmoins}, the antiferromagnetic gap is $2\Delta$ in specific cases where $\mu^{(1)}=\mu^{(2)} = \epsilon_{k_F}= 0$, such as the $2D$ square lattice at half-filling with first-neighbor hopping only. 

\subsection{Consistency between one- and two-particle properties}
\label{sec:sigmag_tpscplus}
Consistency between one- and two-particle properties can be verified through the Galitski-Migdal equation \eref{eq:SigmaExact} which relates the trace of $\Sigma \mathcal{G}$, one-particle quantities, to the double occupancy. We consider this in the Matsubara-frequency and wave vector domain at the second level of approximation of the TPSC+ approach
\begin{equation}
    \Sigma^{(2)}_\sigma(1,\bar{2})\mathcal{G}^{(2)}_\sigma(\bar{2},1^+) = \frac{T}{N}\sum_{\mathbf{k},ik_n}\Sigma^{(2)}_\sigma(\mathbf{k},ik_n)\mathcal{G}^{(2)}_\sigma(\mathbf{k},ik_n).
\end{equation}
Inserting the equation for the self-energy and using $\chi_{sp,ch}(\mathbf{q},iq_n) = \chi_{sp,ch}(-\mathbf{q},-iq_n)$, we obtain
\begin{align}
    &\Sigma^{(2)}_\sigma(1,\bar{2})\mathcal{G}^{(2)}_\sigma(\bar{2},1^+) = \frac{Un^2}{4}\nonumber \\
    &-\frac{UT}{16N}\sum_{\mathbf{q},iq_n}\left [3U_{sp}\chi_{sp}(\mathbf{q},iq_n)+U_{ch}\chi_{ch}(\mathbf{q},iq_n)\right]\chi^{(2)}(\mathbf{q},iq_n),
    \label{eq:traceinter}
\end{align}
which can be solved using the relations
\begin{align}
    \chi_{sp}(\mathbf{q},iq_n) - \chi^{(2)}(\mathbf{q},iq_n) &= \frac{U_{sp}}{2} \chi_{sp}(\mathbf{q},iq_n)\chi^{(2)}(\mathbf{q},iq_n),\\
    \chi^{(2)}(\mathbf{q},iq_n) - \chi_{ch}(\mathbf{q},iq_n)  &= \frac{U_{ch}}{2} \chi_{ch}(\mathbf{q},iq_n)\chi^{(2)}(\mathbf{q},iq_n).
\end{align}
Substituting these relations in \eref{eq:traceinter} and using the sum rules for $\chi_{sp,ch}$ and $\chi^{(2)}$, we find that
\begin{equation}
    \Sigma^{(2)}_\sigma(1,\bar{2})\mathcal{G}^{(2)}_\sigma(\bar{2},1^+) = U\langle n_\uparrow n_\downarrow \rangle,
    \label{eq:trsigmag_tpsc+}
\end{equation}
which is the exact result expected from \eref{eq:SigmaExact}. Hence, at the second level of approximation, the TPSC+ approach shows consistency between single-particle properties such as the self-energy and the Green's function and two-particle properties like the double occupancy. This is another improvement over the original TPSC approach, in which this consistency exists with the trace of $\Sigma^{(2)}\mathcal{G}^{(1)}$ instead of that of $\Sigma^{(2)}\mathcal{G}^{(2)}$. 

\begin{figure}
    \centering
    \includegraphics[width=\columnwidth]{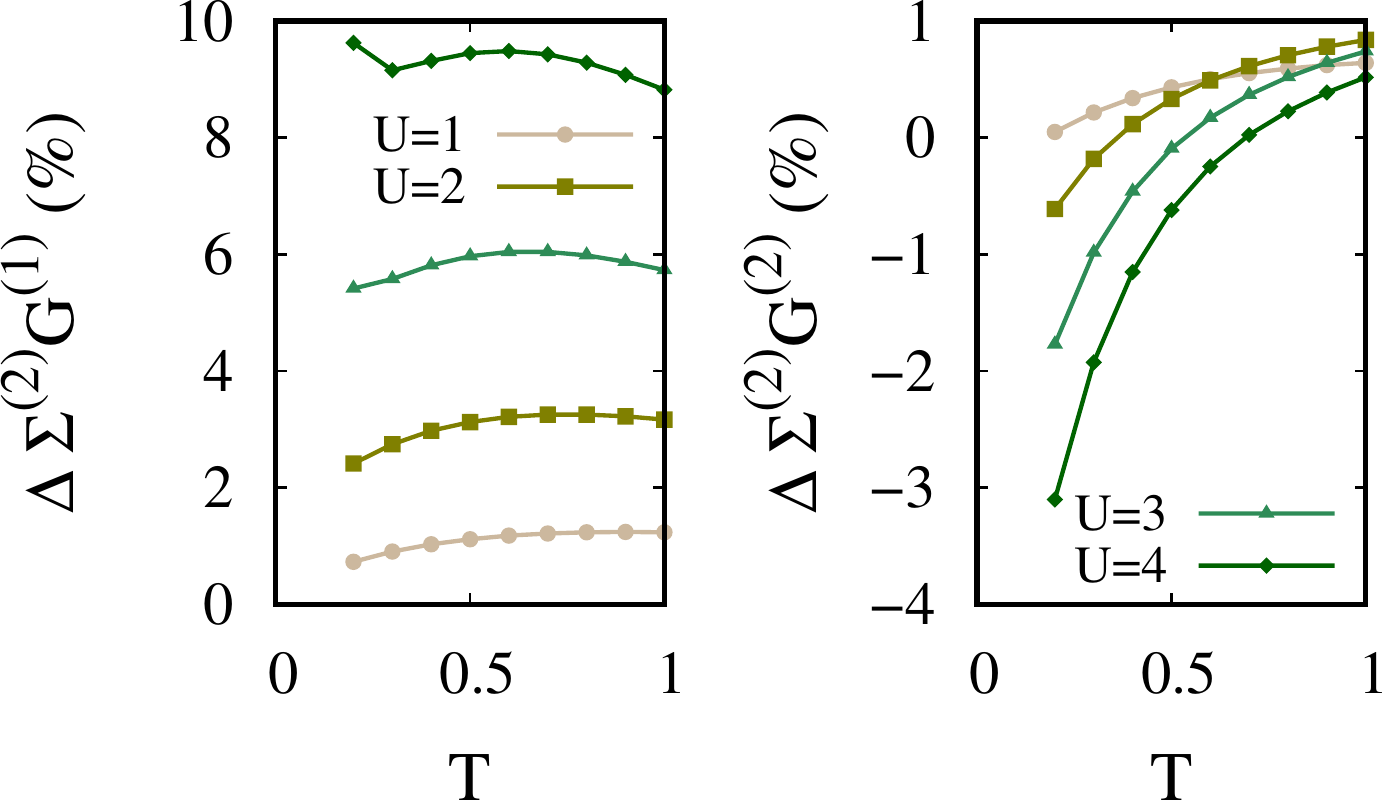}
    \caption{Relative deviation between the traces $\mathrm{Tr}\Sigma^{(2)}\mathcal{G}^{(i)}$ and the expected exact result of \eref{eq:SigmaExact} from the TPSC+SFM approach. On the left, the trace is computed with the non-interacting Green's function $\mathcal{G}^{(1)}$, whereas on the right, it is computed with the interacting Green's function $\mathcal{G}^{(2)}$. The calculations are done for the $2D$ square lattice, at half-filling, with nearest-neighbour hopping only.}
    \label{fig:deviationTraceTPSC+SFM}
\end{figure}

This is not true for the case of the TPSC+SFM approach: the traces of the product of the self-energy and the Green's function, both the non-interacting $\mathcal{G}^{(1)}$ and interacting $\mathcal{G}^{(2)}$, are not expected to yield the exact expected result \eref{eq:SigmaExact}. However, we show in \fref{fig:deviationTraceTPSC+SFM} that the deviation between the expected result and the trace with the interacting Green's function remains of the order of a few percent, and that the deviation is smaller for the trace with the interacting Green's function than with the non-interacting one. 

\subsection{Pseudogap in the $2D$ Hubbard model}
\label{sec:pg_tpscplus}
The pseudogap from antiferromagnetic fluctuations in the weak correlation regime of the Hubbard model has been observed with multiple numerical methods \cite{Schafer_2021}, though it was first predicted by the TPSC approach \cite{Vilk_1997}. Here, we show that the same phenomenology is obtained with the TPSC+ approach.

The antiferromagnetic pseudogap appears in the renormalised classical regime, where the spin fluctuations are dominant. In this regime, we once again use the Ornstein-Zernicke of Eq.(\ref{eq:chispOZ}) for the spin susceptibility and, therefore, the self-energy of Eq.(\ref{eq:selfOZ}).

As described in Refs. \cite{VilkShadow:1997} and \cite{gauvin-ndiaye_disorder_2022}, we evaluate the self-energy at the hot spots $\mathbf{k}_F$ and at zero frequency $\omega=i0^+$. To do so, we change the variables from $\mathbf{q} -\mathbf{Q }\rightarrow \mathbf{q}$ and approximate $\epsilon_{\mathbf{k}_F+\mathbf{q}}$ by $\mathbf{v}_{\mathbf{k}_F} \cdot \mathbf{q}$ with the Fermi velocity at the hot spots connected by $\mathbf{Q}$ to the Fermi wave vector we are interested in. We obtain
\begin{equation}
    \Sigma^R_{cl}(\mathbf{k}_F,0) = \frac{3UT}{4\xi_0^2} \int \frac{d^2q}{(2\pi)^2}\frac{1}{q_{\perp}^2+q_{\parallel}^2+\xi_{sp}^{-2}} \frac{1}{i0^+-q_\parallel v_{F}},
\end{equation}
where the wave vector $\mathbf{q}$ is separated in components parallel ($q_{\parallel}$) and perpendicular ($q_{\perp}$) to the Fermi velocity. The integration can then be performed in the complex plane,~\cite{Vilk_1997,VilkShadow:1997} leading to the following imaginary part of the self-energy

\begin{equation}
    \mathrm{Im}\Sigma_{cl}(\mathbf{k}_F,0) = -\frac{3UT}{16 \xi_0^2}\frac{\xi_{sp}}{\xi_{th}},
    \label{eq:ImSigmaCl}
\end{equation}
with $\xi_{th}=\frac{v_F}{\pi T}$ the  thermal de Broglie wavelength. As a reminder, the spectral weight $A(\mathbf{k},\omega)$ is written as a function of the self-energy
\begin{equation}
    A(\mathbf{k},\omega) = -2\frac{\mathrm{Im}\Sigma(\mathbf{k},\omega)}{(\omega -\epsilon_\mathbf{k}+\mu - \mathrm{Re}\Sigma(\mathbf{k},\omega))^2+(\mathrm{Im}\Sigma(\mathbf{k},\omega))^2}.
\end{equation}
At the Fermi level, if the absolute value of the imaginary part of the self-energy is large, the spectral weight is suppressed. Conversely, a small imaginary part of the self-energy leads to a large value of the spectral weight. Hence, from \eref{eq:ImSigmaCl}, the TPSC+ approach predicts a suppression of the spectral weight $A(\mathbf{k}_F,0)$ when the antiferromagnetic spin correlation length $\xi_{sp}$ becomes larger than the thermal de Broglie wavelength $\xi_{th}$, which is known as the Vilk criterion. This phenomenology corresponds to the appearance of a pseudogap from antiferromagnetic spin fluctuations. Moreover, the same arguments detailed in Ref. \cite{Vilk_1997} can be used to show that two peaks appear at finite frequency in the spectral weight when the Vilk criterion is satisfied.

\begin{figure}
    \centering
    \includegraphics[width=0.7\columnwidth]{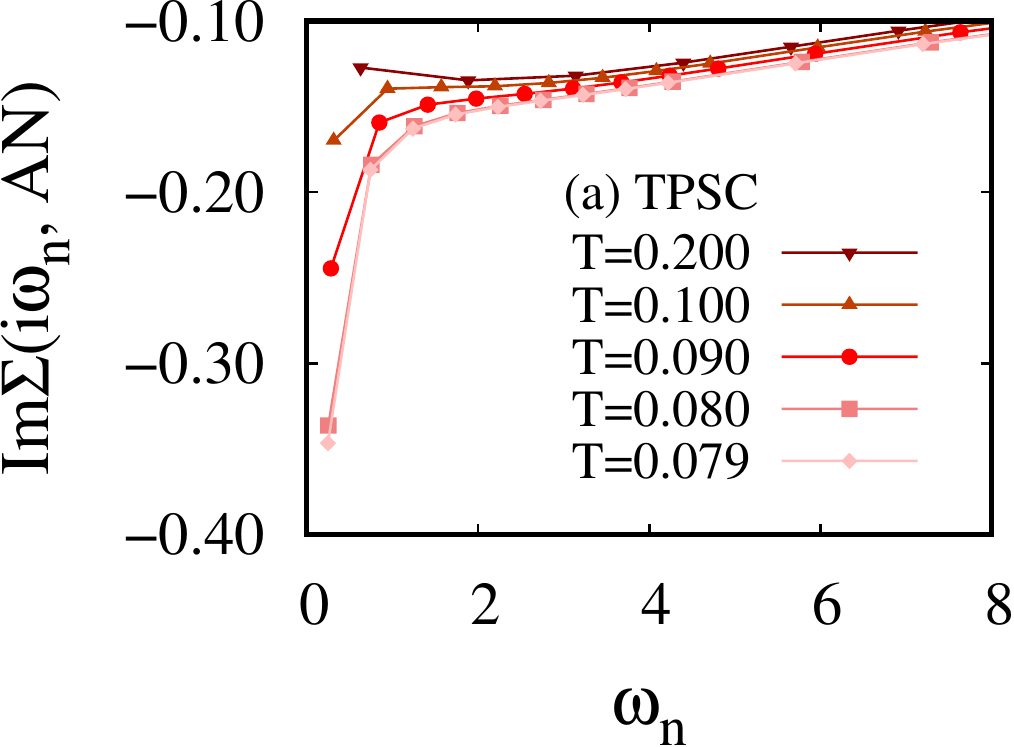}\\
    \includegraphics[width=0.7\columnwidth]{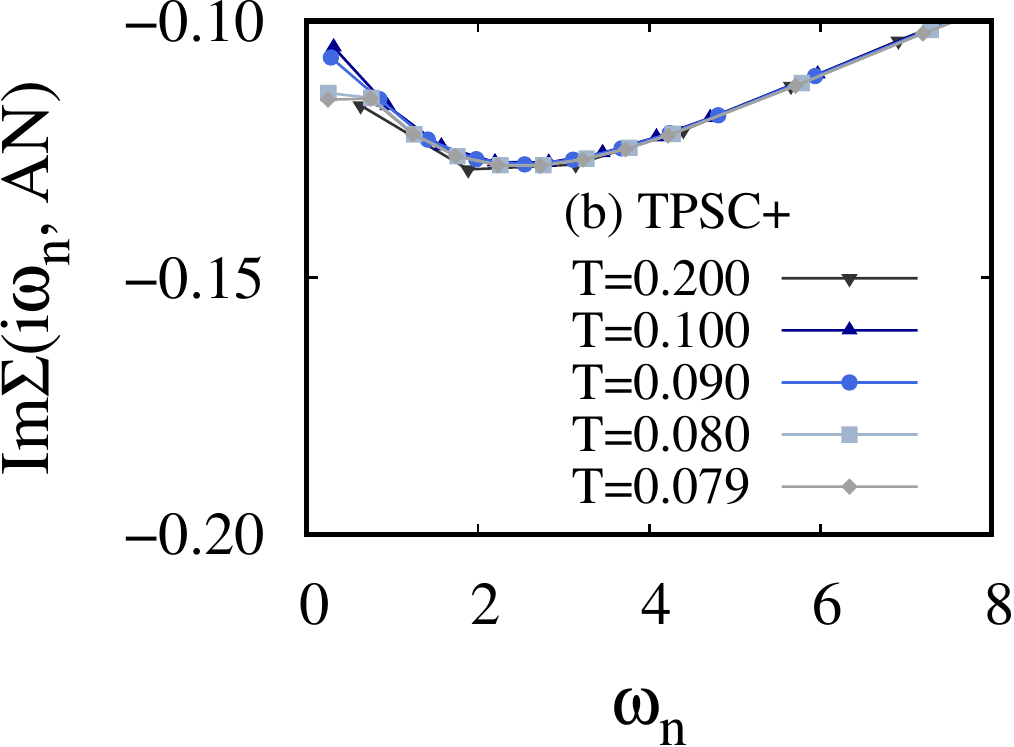}
    \includegraphics[width=0.7\columnwidth]{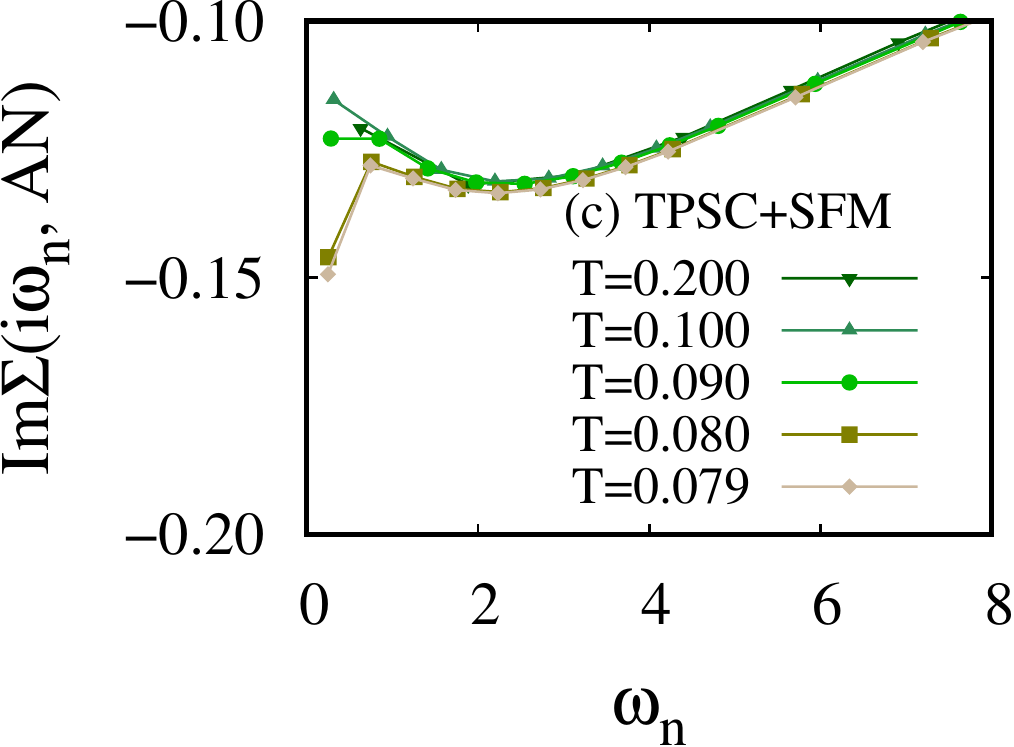}
    \caption{Imaginary part of the self-energy computed with (a) TPSC, (b) TPSC+ and (c) TPSC+SFM at the antinodal point $\mathbf{k}=(\pi,0)$. The calculations are done for the $2D$ square lattice, at half-filling, with nearest-neighbour hopping only, with $U=2$. All three methods predict the opening of a pseudogap as the temperature decreases, as shown by the value of the imaginary part of the self-energy at $\omega_0$ that becomes more negative than the value at $\omega_1$.}
    \label{fig:selfEnergy_AN_PG}
\end{figure}

In \fref{fig:selfEnergy_AN_PG}, we show the imaginary part of the self-energy evaluated at the antinodal point $\mathbf{k}=(\pi,0)$ as a function of the Matsubara frequency for different temperatures, at half-filling and $U=2$. The calculations were performed on the square lattice with nearest-neighbour hopping only. This figure shows numerically that all three TPSC methods can indeed show the opening of a pseudogap at weak-coupling in the $2D$ Hubbard model, though at different temperatures.

\subsection{Limitations: conservation laws and f-sum rule}
\label{sec:lim_tpscplus}
We now turn to some of the limitations of the TPSC+ approach. More specifically, we show that this method does not respect  conservation laws and the f-sum rule.

In the Hubbard model, the f-sum rule for the spin and charge susceptibilities is \cite{Vilk_1997}
\begin{equation}
    \int \frac{d\omega}{\pi}\omega \chi_{sp,ch}''(\mathbf{q},\omega) = \frac{1}{N}\sum_{\mathbf{k},\sigma} \left ( \epsilon_{\mathbf{k}+\mathbf{q}} + \epsilon_{\mathbf{k}-\mathbf{q}} - 2\epsilon_{\mathbf{k}} \right) n^{(2)}_{\mathbf{k},\sigma},
    \label{eq:fsumrule}
\end{equation}
where $n^{(2)}_{\mathbf{k},\sigma}$ is the spin- and momentum-resolved distribution function (the Fermi function in the non-interacting case) computed from the interacting Green's function. We now show that this sum rule is satisfied to some level within the original TPSC approach \cite{Vilk_1997} and the TPSC+SFM modification, but is violated within the TPSC+ approach. 

We start from the spectral representation of the susceptibilities
\begin{align}
    \chi_{sp,ch}(\mathbf{q},iq_n) &= \int \frac{d\omega}{\pi}\frac{ \chi_{sp,ch}''(\mathbf{q},\omega) }{\omega-iq_n},\nonumber \\
    &= \frac{1}{q_n^2} \int \frac{d\omega}{\pi}\frac{\omega \chi_{sp,ch}''(\mathbf{q},\omega)}{1+(\omega/q_n)^2},
    \label{eq:fsum_1}
\end{align}
which, at high frequency, reduces to 
\begin{equation}
    \chi_{sp,ch}(\mathbf{q},iq_n) \simeq \frac{1}{q_n^2} \int \frac{d\omega}{\pi}\omega \chi_{sp,ch}''(\mathbf{q},\omega).
    \label{eq:fsum_high_term}
\end{equation}
Hence, to determine if the TPSC approaches satisfy the f-sum rule, we calculate the coefficients of the $1/q_n^2$ term of the high-frequency expansion of the spin (or charge) susceptibility. Indeed, as seen from \eref{eq:fsum_high_term}, these coefficients correspond to the left hand-side of the f-sum rule \eref{eq:fsumrule}. We denote the coefficients with $\alpha^{(i)}$, where $i$ stands for TPSC, TPSC+ or TPSC+SFM, first recalling that
 \begin{equation}
     \chi_{sp,ch}(\mathbf{q},iq_n) = \frac{\chi^{(i)}(\mathbf{q},iq_n)}{1\mp \frac{U_{sp,ch}}{2}\chi^{(i)}(\mathbf{q},iq_n)}.
     \label{eq:fsum_chisp}
 \end{equation}
 
 Since the non-interacting and partially dressed susceptibilities $\chi^{(i)}$ also behave like $1/q_n^2$ at high frequency, only the numerators of the spin and charge susceptibilities \eref{eq:fsum_chisp} contribute to the coefficients $\alpha^{(i)}$. We now compute the high-frequency expansion of $\chi^{(i)}$ from the spectral representation of the Green's functions. With $A_\sigma^{(i)}$ the spectral weight and $i=1,~2$ denoting the noninteracting and interacting cases respectively, we find 
\begin{align}
    \chi^{(i)}&(\mathbf{q},iq_n) = -\frac{T}{2N}\sum_{k,\sigma} \mathcal{G}_\sigma^{(1)}(k)  \mathcal{G}_\sigma^{(i)}(k+q) + [q\leftrightarrow -q], \nonumber \\
    &= -\frac{T}{2N}\sum_{k,\sigma} \int \frac{d\omega d\omega'}{(2\pi)^2} \frac{A_\sigma^{(1)}(\mathbf{k},\omega)A_\sigma^{(i)}(\mathbf{k}+\mathbf{q},\omega')}{(ik_n -\omega)(ik_n+iq_n-\omega')} \nonumber \\
    &+ [q\leftrightarrow -q], \nonumber \\
    &=  -\frac{1}{N}\sum_{\mathbf{k},\sigma} \int \frac{d\omega d\omega'}{(2\pi)^2} \frac{A_\sigma^{(1)}(\mathbf{k},\omega)A_\sigma^{(i)}(\mathbf{k}+\mathbf{q},\omega')}{q_n^2 + (\omega-\omega')^2}\nonumber \\
    &\times (f(\omega)-f(\omega'))(\omega-\omega') +[\mathbf{q}\leftrightarrow -\mathbf{q}].
\end{align}
From this, we obtain the coefficients $\alpha^{(i)}$
\begin{align}
    \alpha^{(i)} =  &-\frac{1}{2N}\sum_{\mathbf{k},\sigma} \int \frac{d\omega d\omega'}{(2\pi)^2} A_\sigma^{(1)}(\mathbf{k},\omega)A_\sigma^{(i)}(\mathbf{k}+\mathbf{q},\omega')\nonumber \\
    &\times(f(\omega)-f(\omega'))(\omega-\omega') +[\mathbf{q}\leftrightarrow -\mathbf{q}].
\end{align}
The following identities allow us to simplify these results~\cite{nolting1972methode,kalashnikov1973spectral,Vilk_1997}
\begin{align}
   \int \frac{d\omega}{2\pi}A_\sigma^{(i)}(\mathbf{k},\omega) &= 1,\\
    \int \frac{d\omega}{2\pi}A_\sigma^{(i)}(\mathbf{k},\omega)f(\omega) &= n^{(i)}_{\mathbf{k},\sigma},\\
    \int \frac{d\omega}{2\pi}A_\sigma^{(i)}(\mathbf{k},\omega)\omega = \epsilon_\mathbf{k}&-\mu^{(i)} + U\frac{n}{2}\delta_{i,2},\\
   \frac{1}{N} \sum_{\mathbf{k}} \int \frac{d\omega}{2\pi}A_\sigma^{(i)}(\mathbf{k},\omega)\omega f(\omega) &= U\langle n_\uparrow n_\downarrow \rangle \delta_{i,2} \nonumber \\
   + \frac{1}{N} \sum_{\mathbf{k}}&(\epsilon_\mathbf{k}-\mu^{(i)})n_{\mathbf{k},\sigma}^{(i)}. \label{eq:fsr_identity}
\end{align}
Here, $\mu^{(i)}$ is the chemical potential, and $n^{(i)}_{\mathbf{k},\sigma}$ is the spin- and momentum-resolved particle distribution function computed with the non-interacting ($i=1$) or interacting ($i=2$) Green's function. We also note that \eref{eq:fsr_identity} is only valid at level $i=2$ when $\mathrm{Tr}[\Sigma^{(2)}\mathcal{G}^{(2)}]=U\langle n_\uparrow n_\downarrow\rangle$, which is true in the case of TPSC+. From this, we find that the coefficient of the $1/q_n^2$ term is
\begin{align}
    \alpha^{(i)} = &\frac{1}{2N}\sum_{\mathbf{k},\sigma}(\epsilon_{\mathbf{k}+\mathbf{q}}+\epsilon_{\mathbf{k}-\mathbf{q}}-2\epsilon_\mathbf{k})(n^{(1)}_{\mathbf{k},\sigma}+n^{(i)}_{\mathbf{k},\sigma}) \nonumber \\
    &+ U\frac{n^2}{2}\delta_{i,2} - 2U\langle n_\uparrow n_\downarrow \rangle \delta_{i,2}.
\end{align}
For TPSC, with $i=1$, we obtain
\begin{equation}
    \alpha^{(\mathrm{TPSC})} = \frac{1}{N}\sum_{\mathbf{k},\sigma}(\epsilon_{\mathbf{k}+\mathbf{q}}+\epsilon_{\mathbf{k}-\mathbf{q}}-2\epsilon_\mathbf{k})n^{(1)}_{\mathbf{k},\sigma}.
    \label{eq:coeff_tpsc}
\end{equation}
The coefficient for TPSC+SFM is identical to that of TPSC since, at high frequency, the spin and charge susceptibilities are calculated from the non-interacting Lindhard function $\chi^{(1)}$ (see \eref{eq:chi2m}). This term differs from the right-hand side of the f-sum rule \eref{eq:fsumrule} in the particle distribution function: TPSC and TPSC+SFM satisfy the f-sum rule at the non-interacting level, but not at the interacting level.   

For TPSC+, where $i=2$, we find instead
\begin{align}
    \alpha^{(\mathrm{TPSC+})} &= \frac{1}{2N}\sum_{\mathbf{k},\sigma}(\epsilon_{\mathbf{k}+\mathbf{q}}+\epsilon_{\mathbf{k}-\mathbf{q}}-2\epsilon_\mathbf{k})(n^{(1)}_{\mathbf{k},\sigma}+n^{(2)}_{\mathbf{k},\sigma}) \nonumber \\
    &+ U\frac{n^2}{2} - 2U\langle n_\uparrow n_\downarrow \rangle .
    \label{eq:coeff_tpscplus}
\end{align}
Compared with the right-hand side of \eref{eq:fsumrule}, the coefficient for TPSC+ deviates from the expected value of the f-sum rule by the term $U\frac{n^2}{2} - 2U\langle n_\uparrow n_\downarrow \rangle$ at $\mathbf{q}=0$. It is important to note that the f-sum rule is satisfied exactly at $\mathbf{q}=0$ by both TPSC and TPSC+SFM. 

We now assess the deviation of the coefficients \eref{eq:coeff_tpsc} and \eref{eq:coeff_tpscplus} from the f-sum rule numerically. In practice, we evaluate: (a) the coefficients \eref{eq:coeff_tpsc} and \eref{eq:coeff_tpscplus}, (b) the expected value of the f-sum rule from the right-hand side of \eref{eq:fsumrule}, and (c) the left-hand side of \eref{eq:fsumrule}. This last result is obtained from a derivative in imaginary time of the spin susceptibility, 
\begin{align}
    \int \frac{d\omega}{\pi}\omega &\chi_{sp,ch}''(\mathbf{q},\omega) \nonumber \\
    &= \lim_{\eta \rightarrow 0}T\sum_{iq_n}(e^{-iq_n\eta}-e^{iq_n\eta})iq_n\chi_{sp,ch}(\mathbf{q},iq_n),\\
    &= -2\left . \frac{\partial \chi_{sp,ch}(\mathbf{q},\tau)}{\partial \tau}\right |_{\tau=0+},
    \label{eq:derivfsr}
\end{align}
which we compute numerically using finite differences. 

\begin{figure}
    \centering
    \includegraphics[width=0.7\columnwidth]{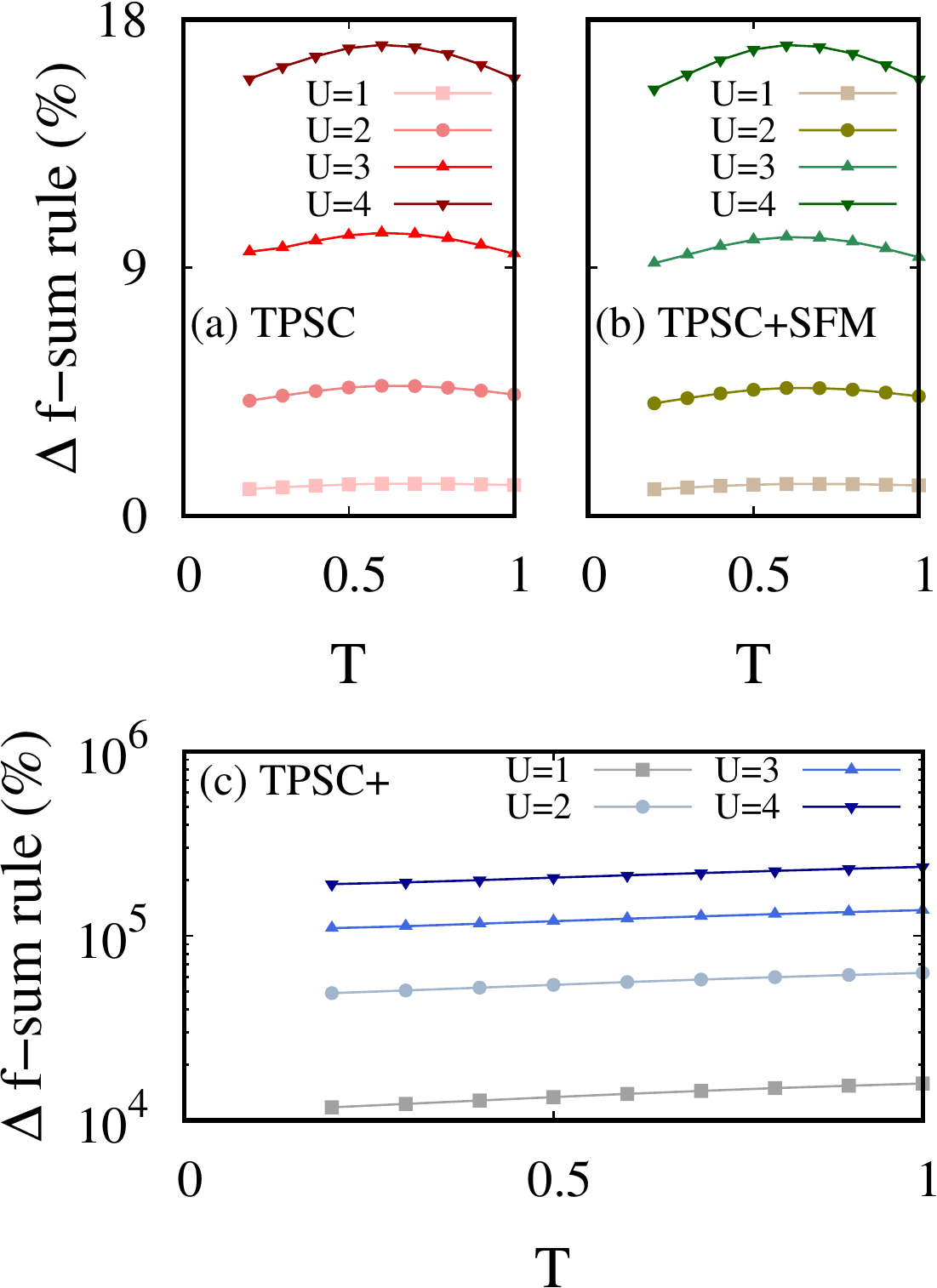}
    \caption{Violation of the f-sum rule evaluated at $\mathbf{q}=(2\pi/N,0)$ with $N=256$ sites in the $x$ direction and density $n=1$ by (a) TPSC, (b) TPSC+SFM and (c) TPSC+ approaches for various values of $U$ as a function of the temperature $T$. More specifically, the plots show the relative deviation between the coefficient computed with the derivatives \eref{eq:derivfsr} and the f-sum rule $\frac{1}{N}\sum_{\mathbf{k},\sigma} \left ( \epsilon_{\mathbf{k}+\mathbf{q}} + \epsilon_{\mathbf{k}-\mathbf{q}} - 2\epsilon_{\mathbf{k}} \right) n^{(2)}_{\mathbf{k},\sigma}$.
    }
    \label{fig:fsum2}
\end{figure}

In \fref{fig:fsum2}, we show the relative deviation between the coefficients computed with \eref{eq:derivfsr} and the expected value of the f-sum rule (right-hand side of \eref{eq:fsumrule}). More specifically, we evaluate this deviation at $\mathbf{q}=(2\pi/N,0)$ with $N=256$ the number of sites in the $x$ direction, since we expect it to be largest at small values of $\mathbf{q}$ where the f-sum rule should be zero. The small relative deviations shown in the panels (a) and (b) for TPSC and TPSC+SFM are due to the fact that both approaches satisfy the f-sum rule at the non-interacting level. In contrast, the large deviation seen in panel (c) for TPSC+ comes from the absolute value of the coefficient at $\mathbf{q}=0$ for this method, $U\frac{n^2}{2} - 2U\langle n_\uparrow n_\downarrow \rangle$, which becomes more important as $U$ increases.

\begin{figure}
    \centering
    \includegraphics[width=0.6\columnwidth]{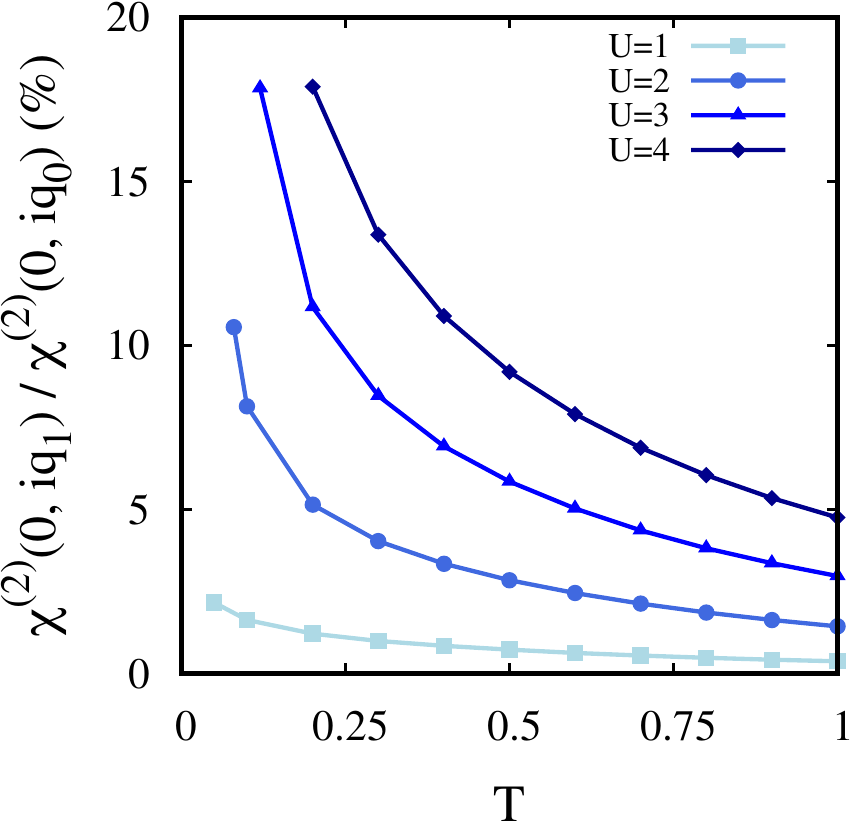} 
    \caption{Ratio of $\chi^{(2)}(\mathbf{q}=0,iq_1)/\chi^{(2)}(\mathbf{q}=0,iq_0)$ as a function of the temperature $T$ obtained from TPSC+ calculations. The TPSC+ approach violates conservation laws, as shown by this non-zero ratio. The calculations are done for the $2D$ square lattice, at half-filling, with nearest-neighbour hopping only.}
    \label{fig:chi2q0w1}
\end{figure}

Another property that follows from spin and charge conservation, is that the spin and charge susceptibilities evaluated at zero wave vector and finite Matsubara frequency should be zero: $\chi_{sp,ch}(\mathbf{q}=0,iq_n\neq 0) =0$. In TPSC, this is achieved because the Lindhard susceptibility is zero at wave-vector $\mathbf{q}=0$ for all non-zero Matsubara frequencies. This is also true in TPSC+SFM. However, it is not the case in TPSC+, where the partially dressed susceptibility $\chi^{(2)}$ remains finite at $\mathbf{q}=0$ for non-zero Matsubara frequencies. We expect the largest deviation to occur at the $n=1$ Matsubara frequency $q_1$. Hence, in \fref{fig:chi2q0w1}, we show for TPSC+ the value of the partially dressed susceptibility $\chi^{(2)}(\mathbf{q}=0,iq_1)$ divided by $\chi^{(2)}(\mathbf{q}=0,iq_0)$, the value at the Matsubara frequency $q_0=0$, as a function of the temperature for $U=1,~2,~3$ and $4$, once again for the $2D$ square lattice at half-filling with nearest-neighbor hopping only. The violation of the conservation laws increases as the temperature decreases and as $U$ increases. Though the value at the $n=1$ Matsubara frequency remains small for $U\leq2$ (one order of magnitude smaller than the value at zero Matsubara frequency), it reaches almost $20\%$ of the value at the $n=0$ Matsubara frequency at low temperatures for $U=3$ and $U=4$. 


\section{Results for the $2D$ Hubbard model}
Now that we have introduced the TPSC+ and the TPSC+SFM approaches and their theoretical basis, we apply them to the $2D$ Hubbard model. The aim of this section is to benchmark the methods by comparing their results to available exact diagrammatic Monte Carlo results and to assess their regime of validity. We first benchmark the spin correlation length in the weak interaction regime at half-filling in \sref{sec:results_weak_n1}. In \sref{sec:results_doped}, we benchmark the spin and charge susceptibilities away from half-filling. We end this section with the benchmark of the self-energy in \sref{sec:benchmark_selfEnergy}, where we add a comparison to the self-energy obtained by second-order perturbation theory. All benchmarks provided here are for the $2D$ Hubbard model with nearest-neighbor hopping only. Our energy units are $t=1$, lattice spacing $a=1$, Plank's constant $\hbar=1$ and Boltzmann's constant $k_B=1$. We do not put the factor $1/2$ for the spin. Details of the implementation may be found in Appendix~\ref{app:implementation}.
\label{sec:results}

\subsection{Spin correlation length at half-filling}
\label{sec:results_weak_n1}

\begin{figure}
    \centering
    \includegraphics[width=\columnwidth]{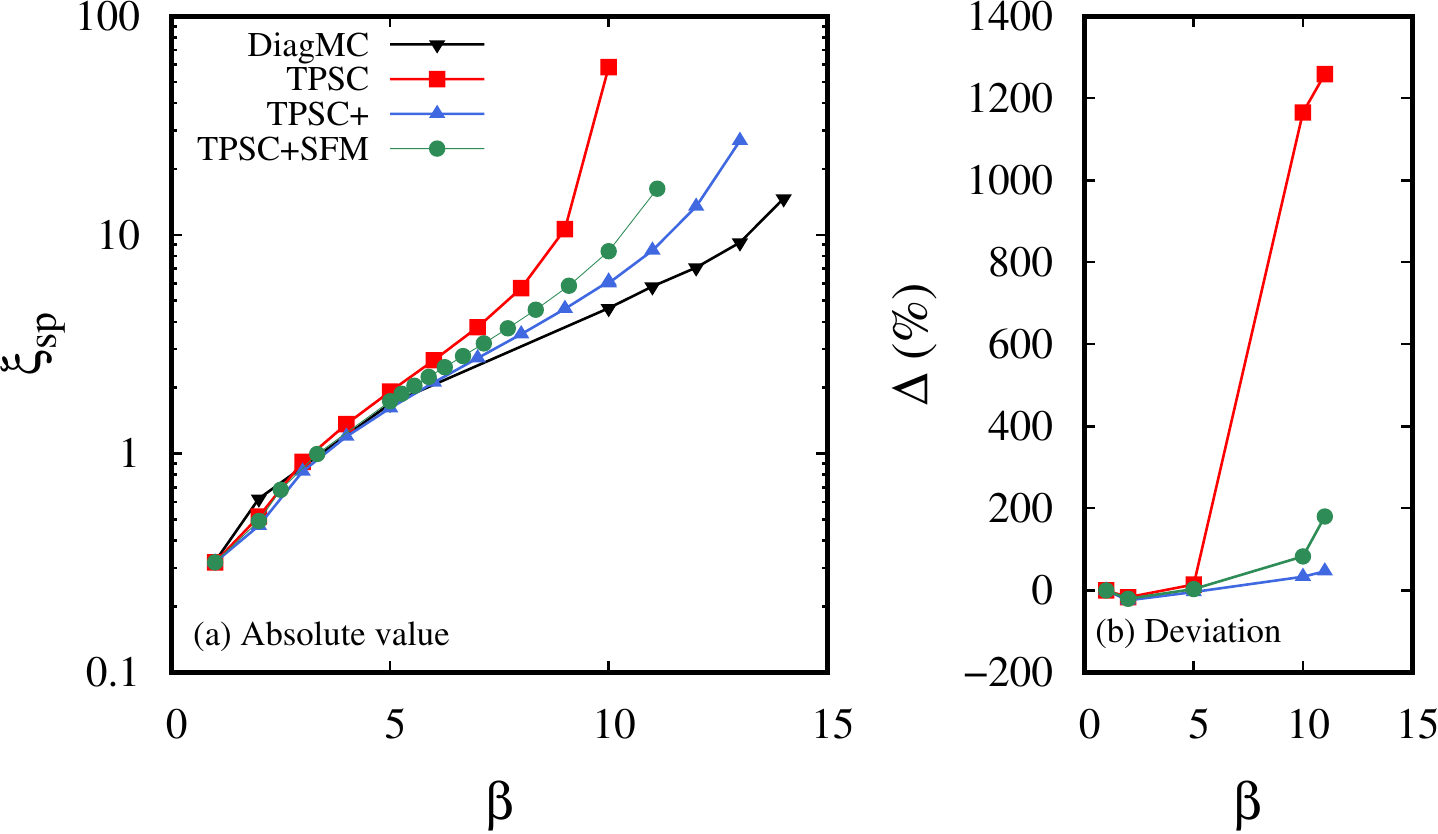} 
    \caption{Spin correlation length from TPSC (red squares), TPSC+ (blue triangles) and TPSC+SFM (green circles) calculations, compared to the DiagMC benchmark (black triangles) from Ref. \cite{Schafer_2021}. The results are obtained as a function of the inverse temperature $\beta$ for the half-filled $2D$ Hubbard model on a square lattice with $U=2$. Panel (a) shows the absolute value of the spin correlation length, while panel (b) shows the relative deviation between the data from all three TPSC methods and the DiagMC benchmark. The relative deviation is calculated as $\Delta = (\xi_{sp,~\mathrm{TPSC}}-\xi_{sp,~\mathrm{DiagMC}})/\xi_{sp,~\mathrm{DiagMC}}$. }
    \label{fig:xisp}
\end{figure}

In \fref{fig:xisp}, we show the spin correlation length obtained for the $2D$ Hubbard at half filling with an interaction strength $U=2$ as a function of the inverse temperature $\beta = 1/T$. 
Results from all three TPSC methods are compared to the DiagMC benchmark data obtained from Ref. \cite{Schafer_2021}. 
Panel (a) shows the absolute value of the spin correlation length, while panel (b) shows the relative deviation between the three TPSC methods and the DiagMC data. 
The relative deviation is calculated as $\Delta = (\xi_{sp,~\mathrm{TPSC}}-\xi_{sp,~\mathrm{DiagMC}})/\xi_{sp,~\mathrm{DiagMC}}$.
We first note that all three TPSC methods yield accurate results at high temperatures ($\beta \lesssim 7$), at a much lower computational cost than DiagMC calculations.
As was shown in Ref. \cite{Schafer_2021}, the spin correlation length obtained from the original TPSC approach deviates strongly from the benchmark data as the temperature decreases. 
In TPSC, this large deviation is due to the entry in the renormalized classical regime around $T=0.1$, below which the method is not valid anymore. 
Both the TPSC+ and TPSC+SFM approach offer a quantitative and qualitative improvement over the TPSC approach in the low-temperature regime of the weakly interacting $2D$ Hubbard model.
Quantitatively, the relative deviations with DiagMC reach $1165\%$ for TPSC, $33\%$ for TPSC+ and $82\%$ for TPSC+SFM at $\beta=10$. 

\subsection{Weak to intermediate interaction regimes away from half-filling}
\label{sec:results_doped}

\subsubsection{Double occupancy}
\begin{figure}
    \centering
    \includegraphics[width=\columnwidth]{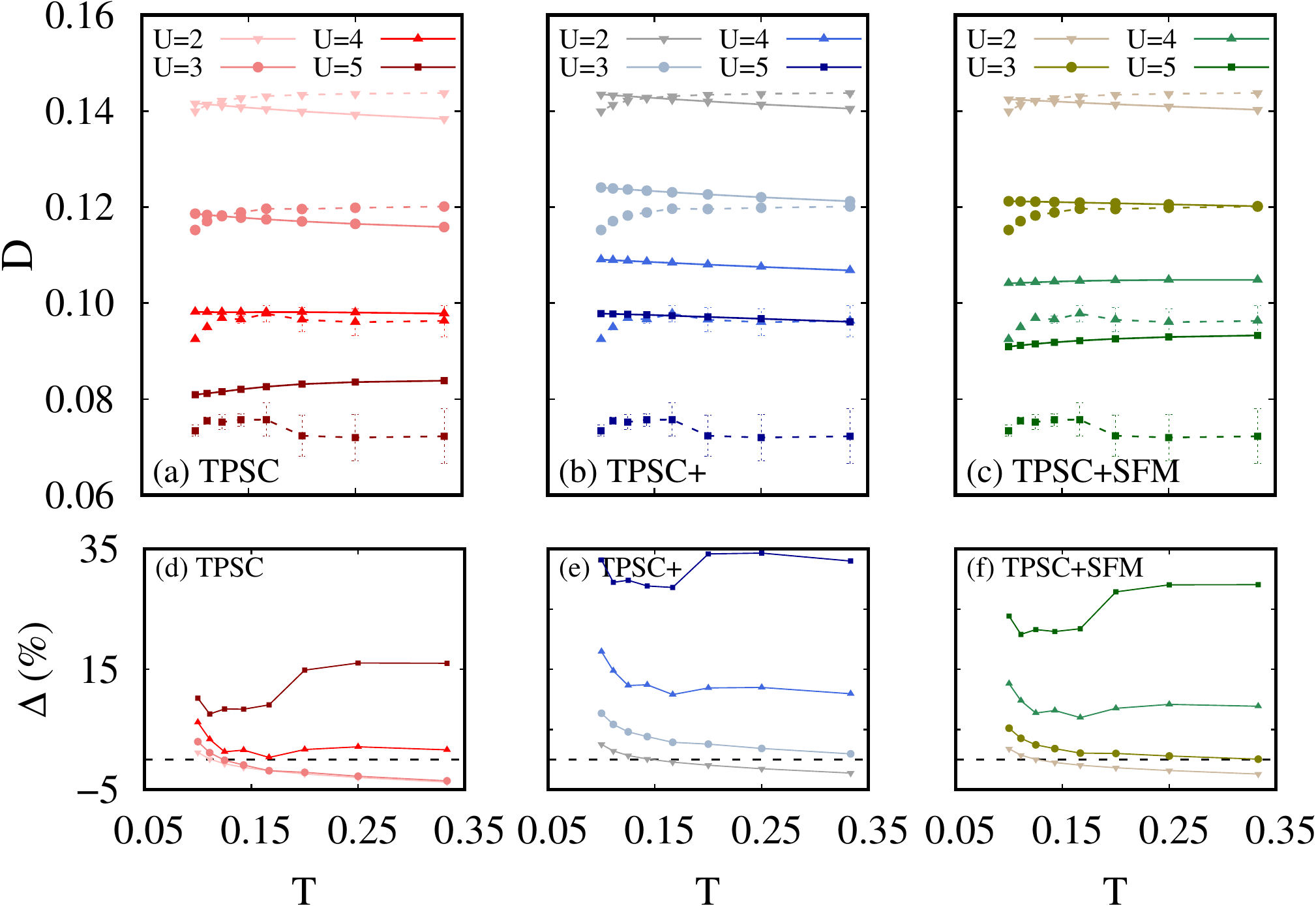} 
    \caption{Top panel: Double occupancy as a function of $T$ and $U$ at fixed density $n=0.875$. Full lines are obtained with (a) TPSC, (b) TPSC+ and (c) TPSC+SFM calculations. CDet data, shown as dashed lines, come from Ref. \cite{Simkovic_2021}. Bottom panel: Relative deviation between the CDet benchmark data and (d) TPSC, (f) TPSC+ and (g) TPSC+SFM data, shown as $\Delta = (D_{\mathrm{TPSC}}-D_{\mathrm{CDet}})/D_{\mathrm{CDet}}$.}
    \label{fig:docc_simkovic}
\end{figure}

\fref{fig:docc_simkovic} shows the temperature and $U$ dependence of the double occupancy computed with (a) TPSC, (b) TPSC+ and (c) TPSC+SFM at fixed density $n=0.875$. We compare our results with CDet benchmark data \cite{Simkovic_2021}. The bottom panel of \fref{fig:docc_simkovic} shows the relative deviation between (d) TPSC, (e) TPSC+ and (f) TPSC+SFM with respect to the CDet benchmark. The relative deviation is calculated as $\Delta = (D_{\mathrm{TPSC}}-D_{\mathrm{CDet}})/D_{\mathrm{CDet}}$.
In the weak interaction regime ($U\leq3$), all three TPSC method yield quantitatively accurate double occupancies: the relative deviations with respect to the CDet benchmark reach at most $5\%$ (in absolute value) at all temperatures. For the higher values of $U$ considered ($U=4$ and $U=5$), the TPSC results are more accurate than the TPSC+ and TPSC+SFM ones. The deviations obtained from TPSC do not exceed $15\%$, while they reach almost $35\%$ for TPSC+ and TPSC+SFM for $U=5$. Qualitatively, the double occupancy should decrease slightly as the temperature is lowered, as seen from the CDet data. This behavior is captured qualitatively at $U=4$ and $U=5$ by TPSC+SFM, but not by TPSC+.

\subsubsection{Spin susceptibility}
\label{sec:benchmark_chisp}
\begin{figure}
    \centering
    \includegraphics[width=\columnwidth]{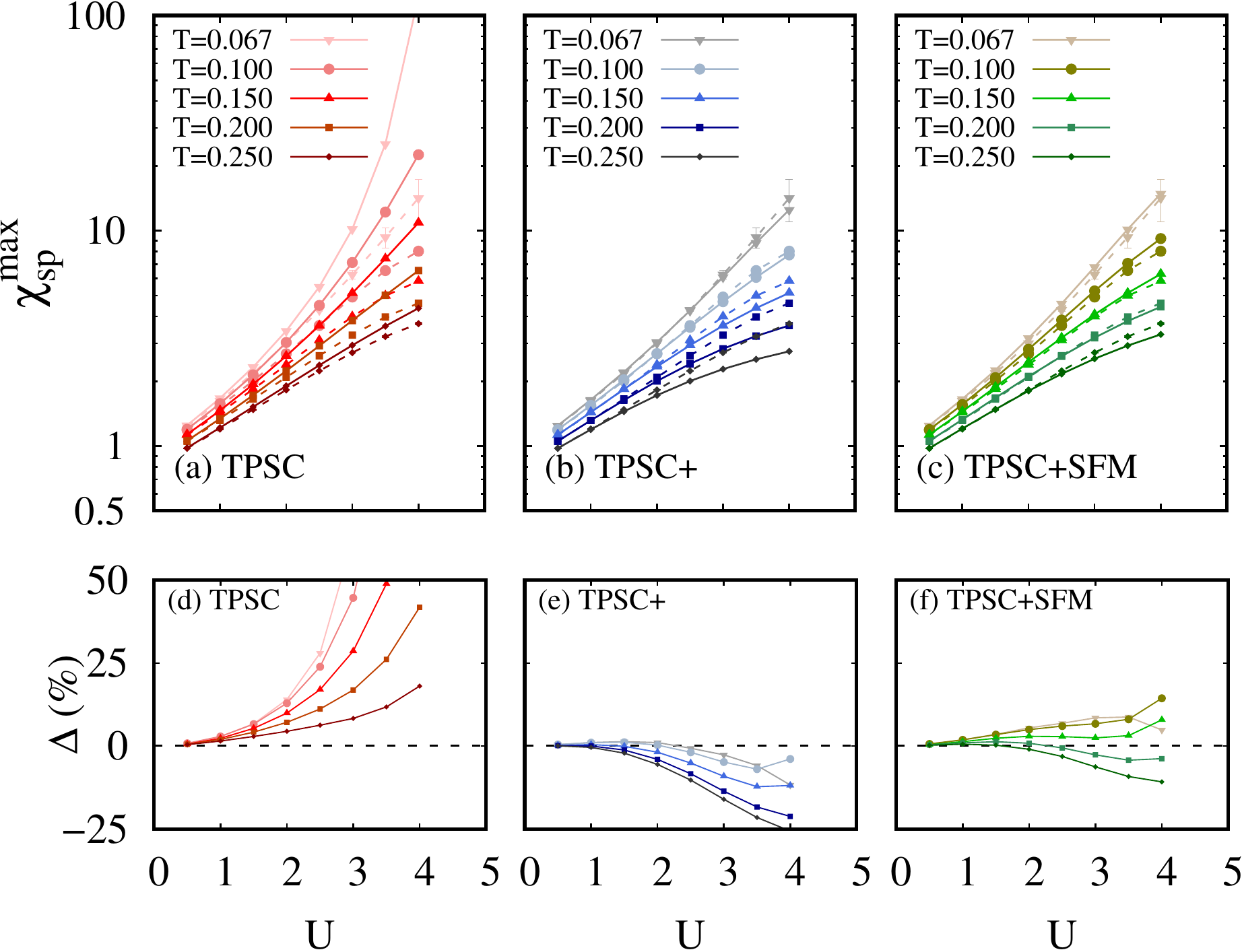} 
    \caption{Top panel: Semi-logarithmic plots of the maximal value of the spin susceptibility as a function of $U$ for five temperatures at fixed density $n=0.875$. Results are shown for (a) TPSC, (b) TPSC+ and (c) TPSC+SFM calculations in full lines. The dotted dashed lines are the CDet data from Ref. \cite{Simkovic_2021}. Bottom panel: Relative deviation between the CDet benchmark data and (d) TPSC, (e) TPSC+ and (f) TPSC+SFM results. }
    \label{fig:chispmax_simkovic}
\end{figure}

We now study the spin susceptibility away from half filling. We first illustrate the maximal value of the spin susceptibility at fixed density $n=0.875$ as a function of $U$ in the weak to intermediate interaction regime ($U\leq 5$) in \fref{fig:chispmax_simkovic} from (a) TPSC, (b) TPSC+ and (c) TPSC+SFM, compared with the CDet benchmark data \cite{Simkovic_2021}. In Appendix \ref{app:strongU}, we show extended results in the strong interaction regime (up to $U=8$), above the validity regime of all three TPSC approaches.

In the bottom panel of \fref{fig:chispmax_simkovic}, we show the relative deviation between (d) TPSC, (e) TPSC+ and (f) TPSC+SFM and the CDet benchmark data, calculated as $\Delta = (\chi^{\mathrm{max}}_{sp,~\mathrm{TPSC}}-\chi^{\mathrm{max}}_{sp,~\mathrm{CDet}})/\chi^{\mathrm{max}}_{sp,~\mathrm{CDet}}$.

The results of all three TPSC variations are in qualitative and quantitative agreement with the exact CDet results in the weakly interacting regime ($U\leq2$), where the relative deviations with respect to the benchmark are below $10\%$ (in absolute value). The deviations increase with the interaction $U$ for all three TPSC methods. 
The TPSC+ and TPSC+SFM approaches offer a significant qualitative and quantitative improvement over the original TPSC approach at low temperatures ($T\leq0.2$). In this temperature regime, the deviations between the TPSC+ and TPSC+SFM data and the CDet benchmark is at most $25\%$, whereas it exceeds $50\%$ with TPSC. 
While TPSC+ is accurate at low temperatures, its deviation with the benchmark increases with temperature. 
The results from TPSC+SFM have the best overall qualitative and quantitative agreement with the benchmark data. In contrast, the maximal value of the spin susceptibility is systematically overestimated by TPSC and underestimated by TPSC+. 

\begin{figure}
    \centering
    \includegraphics[width=\columnwidth]{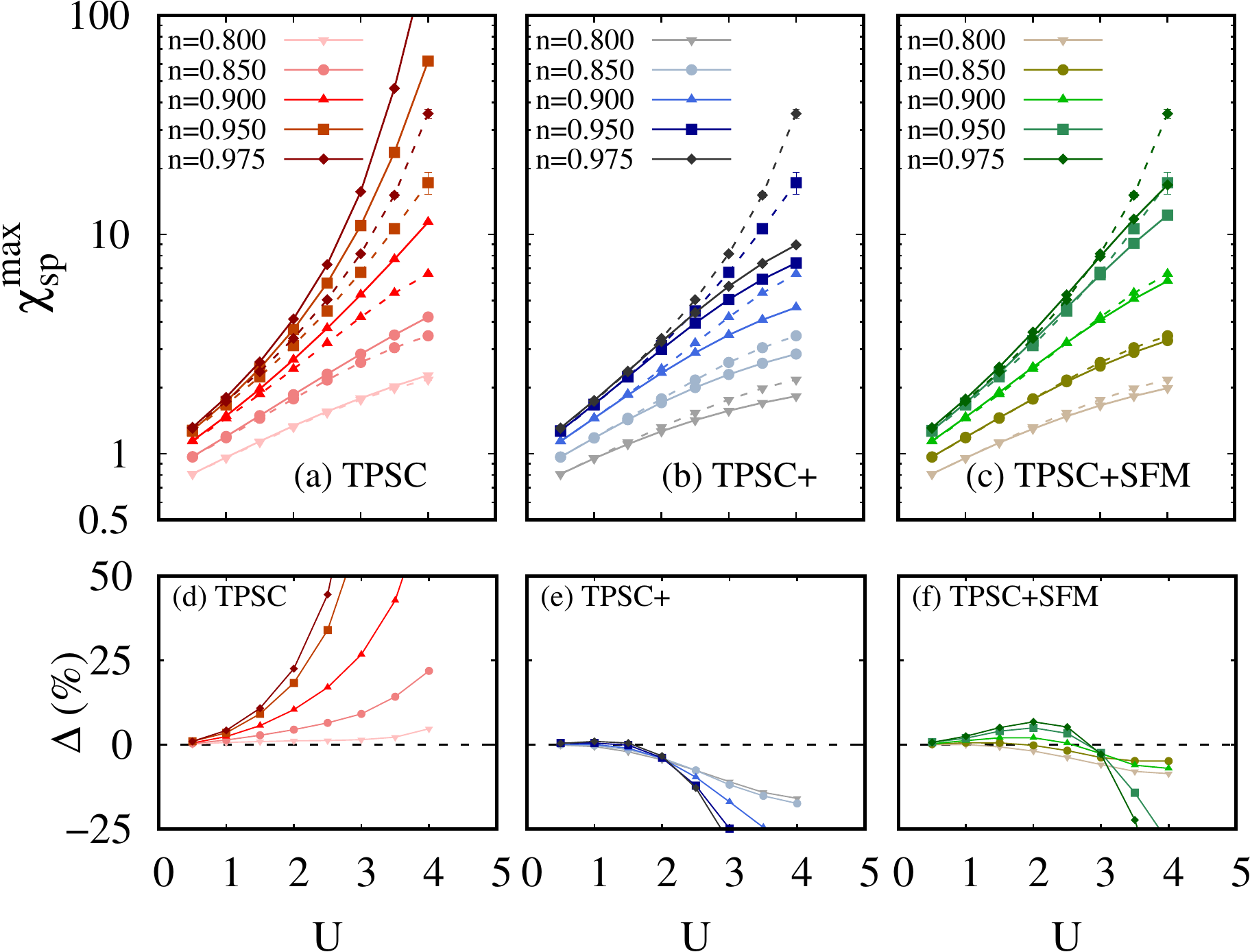}
    \caption{Top panel: Semi-logarithmic plots of the maximal value of the spin susceptibility as a function of $U$ for five different values of filling $n$ at fixed temperature $T=0.2$. Results are shown from (a) TPSC, (b) TPSC+ and (c) TPSC+SFM calculations in full lines. The dotted dashed lines are the CDet data from Ref. \cite{Simkovic_2021}. Bottom panel: Relative deviation between the CDet benchmark data and (d) TPSC, (e) TPSC+ and (f) TPSC+SFM results.}
    \label{fig:chispmax_simkovic_Tfixe}
\end{figure}

In \fref{fig:chispmax_simkovic_Tfixe}, we show the maximal value of the spin susceptibility as a function of the density $n$ and of $U$ at fixed temperature $T=0.2$. 
We first discuss the absolute TPSC results shown in panel (a) as well as their relative deviations with respect to the CDet benchmark shown in panel (d). The TPSC results for the maximal value of the spin susceptibility become more accurate as the density decreases, moving away from half-filling. More specifically, the relative deviation is below $5\%$ at $n=0.8$. This better agreement at low density is due to the the Hartree decoupling used for the TPSC ansatz, which works best in the dilute limit. In contrast, when the density is closer to half filling ($n=1$), TPSC shows deviations that can exceed $25\%$ even in the weakly interacting regime ($U=2$). 
We now turn to the TPSC+ and TPSC+SFM approach results shown in panels (b) and (c) of \fref{fig:chispmax_simkovic_Tfixe} respectively, while panels (d) and (e) show their deviations to the benchmark data. Both TPSC+ and TPSC+SFM tend to underestimate the maximal value of the spin susceptibility for these model parameters. Overall, TPSC+SFM offers the best improvement over the original TPSC approach. TPSC+SFM yields accurate results for low values of $U$ ($U\leq 3$), with deviations to the CDet data that are below $10\%$ in absolute value. The same is true of the TPSC+ results at slightly lower values of $U$ ($U\leq 2$). 

In summary, we conclude from both \fref{fig:chispmax_simkovic} and \fref{fig:chispmax_simkovic_Tfixe} that the maximal value of the spin susceptibility obtained from the TPSC+SFM approach is qualitatively and reasonably quantitatively accurate in the weak to intermediate coupling regime of the $2D$ Hubbard model, away from half filling. In contrast, the TPSC results are accurate in the dilute limit and at high temperatures, while the TPSC+ results are accurate at low temperatures, away from half filling.


\subsubsection{Charge susceptibility}
\label{sec:benchmark_chich}

\begin{figure}
    \centering
    \includegraphics[width=1\columnwidth]{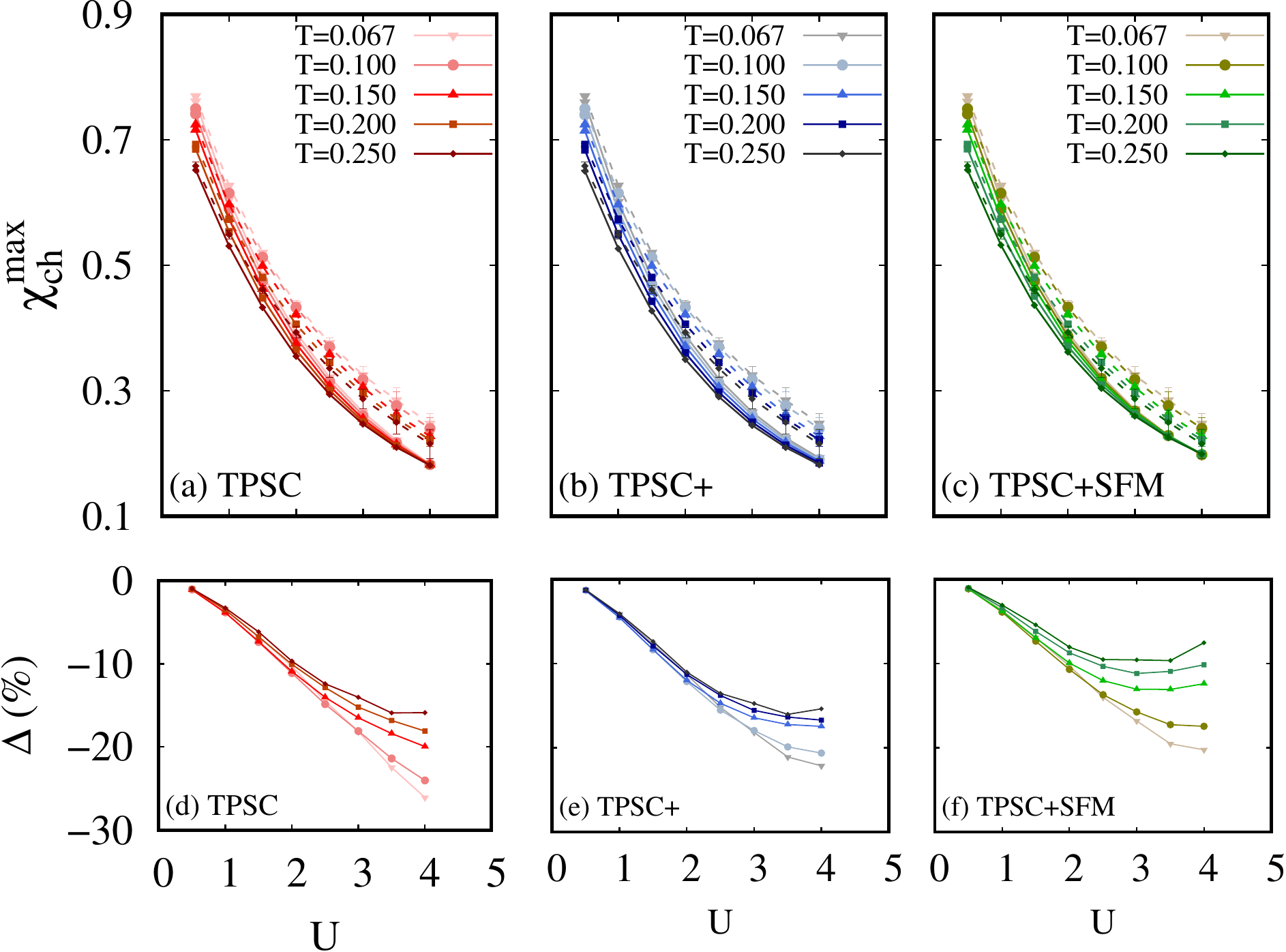}
    \caption{Top panel: Semi-logarithmic plots of the maximal value of the charge susceptibility as a function of $U$ for five different temperatures at fixed filling $n=0.875$. Results are shown for (a) TPSC, (b) TPSC+ and (c) TPSC+SFM calculations in full lines. The dashed lines are the CDet data from Ref. \cite{Simkovic_2021}. Bottom panel: Relative deviation between the CDet benchmark data and the (d) TPSC, (e) TPSC+ and (f) TPSC+SFM results.}
    \label{fig:chichmax_simkovic}
\end{figure}

In \fref{fig:chichmax_simkovic}, we show the maximal value of the charge susceptibility as a function of the temperature and of $U$ at fixed density $n=0.875$. The results from all three TPSC methods are in qualitative agreement with the CDet benchmark data. Quantitatively, the charge susceptibility is underestimated by all three TPSC approaches for all the parameters considered here. The deviations with respect to the CDet benchmark increase with $U$ and as the temperature decreases for the three methods. The TPSC results have the strongest deviations (about $30\%$ at the lowest temperature considered, $T=0.067$, and $U=5$), while the TPSC+SFM results have the best overall agreement with the benchmark (about $20\%$ at most).


\subsection{Benchmark of the self-energy}
\label{sec:benchmark_selfEnergy}

\begin{figure}
    \centering
    \includegraphics[width=1\columnwidth]{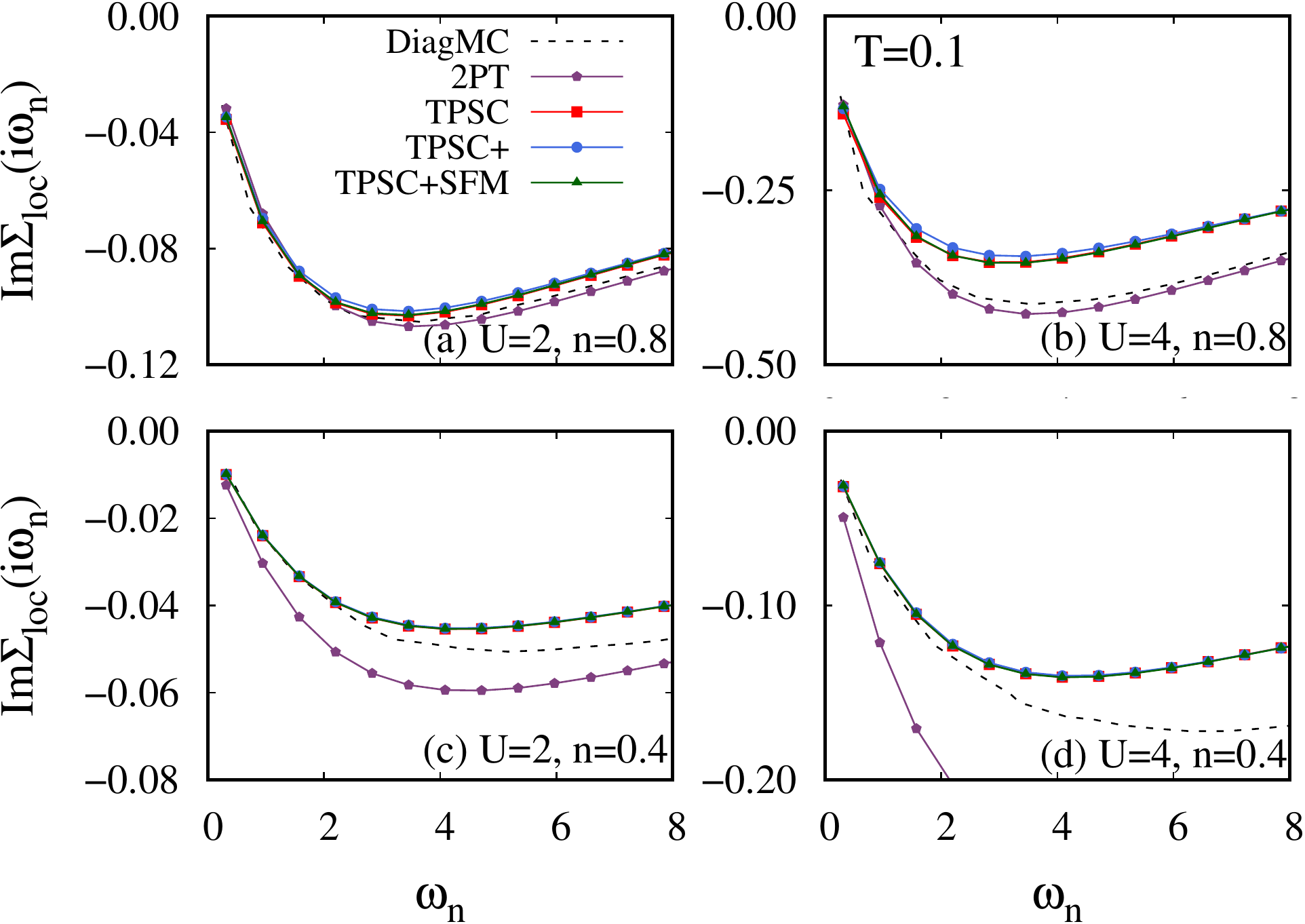}
    \caption{Imaginary part of the local self-energy as a function of the Matsubara frequencies at fixed temperature $T=0.1$, away from half-filling. Calculations are done for (a) $U=2$ and $n=0.8$, (b) $U=4$ and $n=0.8$, (c) $U=2$ and $n=0.4$, and (d) $U=4$ and $n=0.4$. Results are shown for TPSC (red squares), TPSC+ (blue circles), TPSC+SFM (green triangles) and 2PT (purple pentagons) calculations. The benchmark data, in black dashed lines, is from Ref. \cite{Gukelberger_2015}. 
    }
    \label{fig:comp_gukel_v2}
\end{figure}

In \fref{fig:comp_gukel_v2}, we show the imaginary part of the local self-energy as a function of the Matsubara frequencies for $T=0.1$ in the dilute limit ($n=0.4$ and $n=0.8$), and for the Hubbard interaction strengths $U=2$ and $U=4$. The results are obtained from TPSC, TPSC+ and TPSC+SFM calculations. We compare them to DiagMC benchmark data \cite{Gukelberger_2015} and to the self-energy $\Sigma_{2PT}$ obtained from second-order perturbation theory (2PT)
\begin{equation}
    \Sigma_{2PT}(\mathbf{r},\tau) = U^2\mathcal{G}^{(0)}(\mathbf{r},\tau)\mathcal{G}^{(0)}(\mathbf{r},\tau)\mathcal{G}^{(0)}(-\mathbf{r},-\tau).
\end{equation}

For these model parameters, there is no significant difference between the TPSC, TPSC+ and TPSC+SFM results. However, they differ significantly from the 2PT self-energy even in the low interaction case with density $n=0.4$ and interaction $U=2$. This highlights that the three TPSC approaches are non-trivial in their construction.
The results obtained by the three TPSC approaches are accurate at low Matsubara frequencies. These methods can hence properly describe the Fermi liquid properties of the quasiparticles for these model parameters. This is in contrast with 2PT, which is inaccurate at low frequencies even in the dilute limit $n=0.4$. In the latter case it reaches the correct high-frequency tail of the benchmark ~\cite{Gukelberger_2015} at much higher frequency than the range shown in \fref{fig:comp_gukel_v2}. The high-frequency tail of the local part of the self-energy of all three TPSC methods deviates from the benchmark data. This is more pronounced in the $U=4$ case than for $U=2$.
The reasons for the behavior of TPSC at high frequencies is discussed in Appendix E of Ref~\cite{Vilk_1997}. Similar considerations apply to the other versions of TPSC.  

\begin{figure}
    \centering
    \includegraphics[width=1\columnwidth]{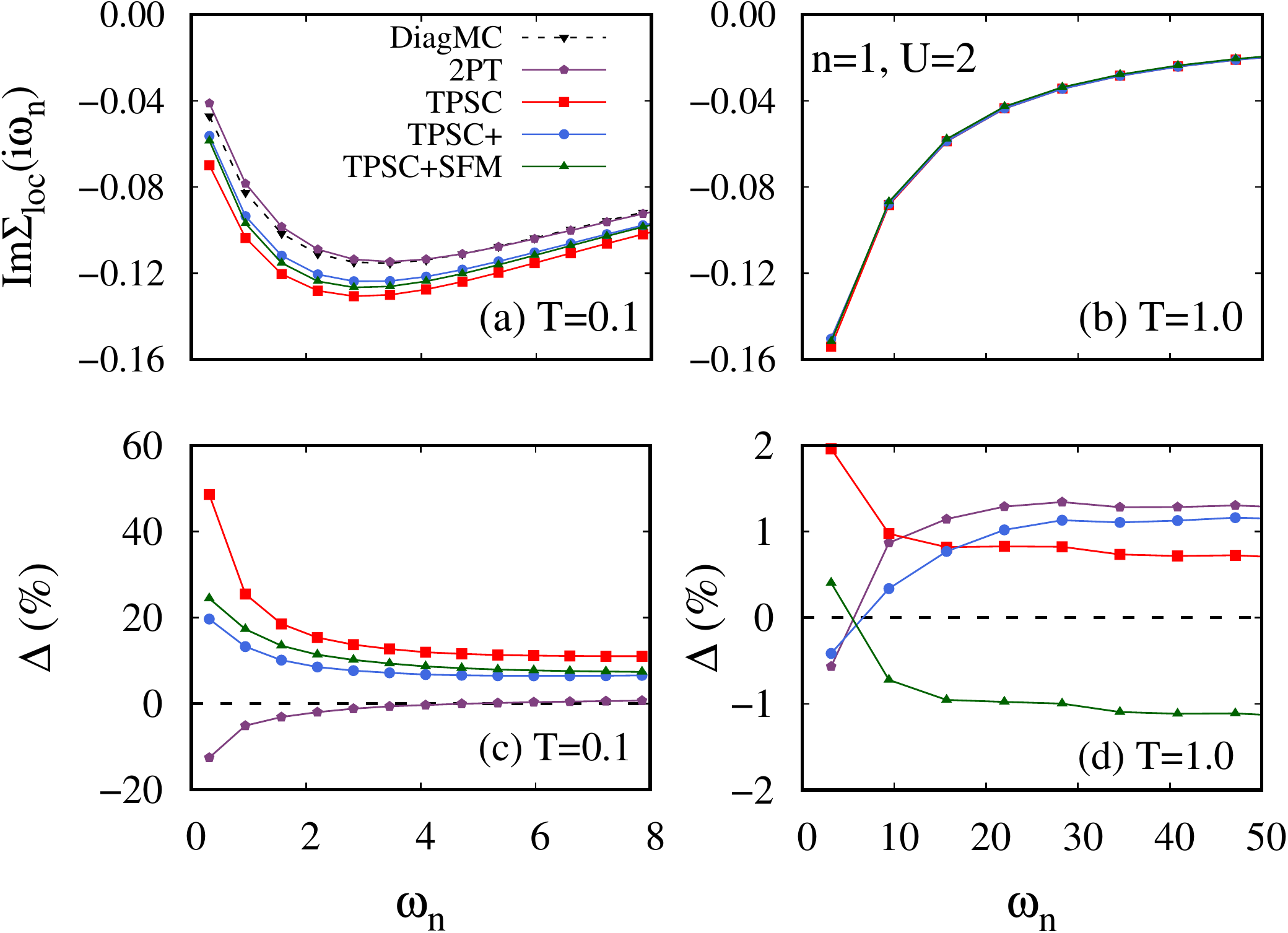}
    \caption{Top panel: Imaginary part of the local self-energy as a function of the Matsubara frequencies at half-filling $n=1$, and $U=2$. Calculations are done for temperatures (a) T=0.1 and (b) T=1.0. The plots show the data from TPSC (red squares), TPSC+ (blue circles), TPSC+SFM (green triangles), and 2PT (purple pentagons) calculations. The benchmark data, in dashed black lines, is obtained from Ref. \cite{Schafer_2021}. Bottom panel: relative deviation between the TPSC, TPSC+, TPSC+SFM and 2PT calculations and the benchmark at temperatures (c) T=0.1 and (d) T=1.0. }
    \label{fig:loc_self_Schafer_v2}
\end{figure}

In panels (a) and (b) of \fref{fig:loc_self_Schafer_v2}, we show the imaginary part of the local self-energy as a function of the Matsubara frequencies at half filling and with a Hubbard interaction strength of $U=2$, for $T=0.1$ and $T=1$. We once again compare the three TPSC methods to the 2PT and benchmark \cite{Schafer_2021} self-energies. Panels (c) and (d) of \fref{fig:loc_self_Schafer_v2} show the relative deviation between all TPSC results and the DiagMC benchmark data.

All the TPSC methods have a good global behaviour. Their results at $T=1$ are accurate, with a deviation to the DiagMC benchmark of at most $2\%$. 
Whereas the TPSC+SFM approach yields the most accurate results for the spin and charge susceptibilities, as shown in \sref{sec:benchmark_chisp} and \sref{sec:benchmark_chich}, the $T=0.1$ results shown in panels (a) and (c) of \fref{fig:loc_self_Schafer_v2} show that the TPSC+ results for the local self-energy are the most accurate ones. 
Similar to the dilute case presented in \fref{fig:comp_gukel_v2}, the high-frequency tail of the self-energy computed with all three TPSC methods is less accurate than that obtained with the second-order perturbation theory. 
Both TPSC+ and TPSC+SFM offer improved results over the original TPSC method at $T=0.1$, with deviations to the DiagMC benchmark of the order of $20\%$ at the lowest Matsubara frequency. In contrast, this deviation reaches about $50\%$ with TPSC. 

\begin{figure}
    \centering
    \includegraphics[width=1\columnwidth]{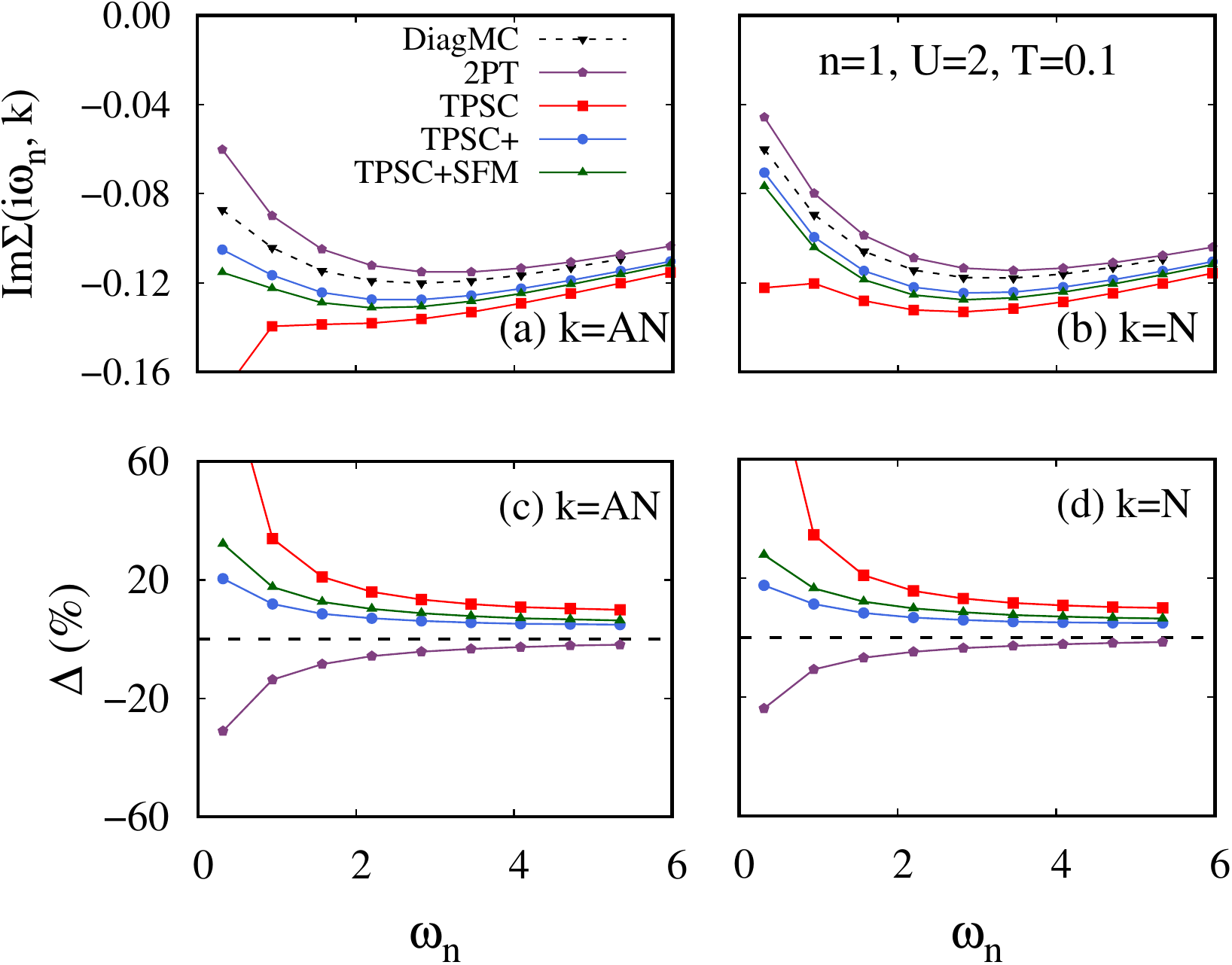}
    \caption{Top panel: Imaginary part of the self-energy as a function of the Matsubara frequencies at half-filling $n=1$, at temperature $T=0.1$ and interaction $U=2$, evaluated at (a) the antinode $\mathbf{k}=(\pi,0)$ and (b) the node $\mathbf{k}=(\pi/2,\pi/2)$. The plots show the data from TPSC (red squares), TPSC+ (blue circles), TPSC+SFM (green triangles), and 2PT (purple pentagons) calculations. DiagMC data, in dashed black lines, is obtained from Ref. \cite{Schafer_2021}. Bottom panel: relative deviation between the TPSC, TPSC+ TPSC+SFM, and 2PT calculations and the DiagMC benchmark for (c) the antinode and (d) the node. }
    \label{fig:self_energy_AN_vs_N}
\end{figure}

In \fref{fig:self_energy_AN_vs_N}, we focus on the low-temperature $T=0.1$ case and show the imaginary part of the self-energy as a function of the Matsubara frequencies at two different wave vectors: the nodal point $\mathbf{k}=(\pi/2,\pi/2)$, and the antinodal point $\mathbf{k}=(\pi,0)$. These results are obtained at half filling and with a Hubbard interaction strength $U=2$. In this model, DiagMC calculations show that the antiferromagnetic pseudogap should open at the antinode at the temperature $T^*_{AN}=0.065$, and at the node at the temperature $T^*_{N}=0.0625$ \cite{Schafer_2021}.
We first note that the 2PT self-energy is closer to that of Fermi liquid quasiparticles than the DiagMC exact results at both $\mathbf{k}$-points. The absolute value of the resulting deviation between the 2PT and DiagMC self-energies is of the order of $20-30\%$ for the first Matsubara frequency. 
In contrast, all three TPSC approaches a self-energy with a smaller quasiparticle weight than the DiagMC results for both $\mathbf{k}$-points. 
The TPSC approach overestimates the temperatures $T^*_{AN}$ and $T^*_N$ at which the pseudogap opens at the antinode and at the node, respectively. This is seen in panels (a) and (b) of \fref{fig:self_energy_AN_vs_N} by the value of the imaginary part of the self-energy at the first Matsubara self-energy $\omega_0$, which is more negative that that at the second Matsubara frequency $\omega_1$. This leads to deviations between TPSC and DiagMC that exceed $60\%$ at the first Matsubara frequency.
As was anticipated in \fref{fig:selfEnergy_AN_PG}, TPSC+ and TPSC+SFM also overestimate the pseudogap temperatures, but in a less pronounced way than TPSC. At $T=0.1$, the deviations between the TPSC+(SFM) and DiagMC self-energies is of the order of $20\%$ ($35\%$). This is a significant improvement over the original TPSC approach.
From this, we conclude that TPSC+ is better suited at describing the self-energy than TPSC+SFM, but TPSC+SFM gives a more accurate description of the spin and charge susceptibilities. 

\subsection{Summary}
In this section, we benchmarked the TPSC, TPSC+ and TPSC+SFM methods against available exact diagrammatic Monte Carlo results for the $2D$ Hubbard model. We showed that, out of the three TPSC methods, the TPSC+ approachs yields the most accurate self-energy results, while the TPSC+SFM approach is the best at describing the spin and charge susceptibilities. By construction, the TPSC+ and TPSC+SFM approaches are not valid in the strongly interacting Hubbard model, just like TPSC. However, they extend the domain of validity of TPSC in the weak to intermediate correlation regime since they are valid at low temperatures in the renormalized classical regime of the $2D$ Hubbard model. Future work will quantify this statement.

From our benchmark work, we conclude that all three TPSC approaches can be reliably used at high temperatures and low densities. At low temperatures and near half filling, the TPSC+ and TPSC+SFM approaches are more accurate than the original TPSC approach.


\section{Conclusion}
\label{sec:conclusion}

In this work, we introduced two improved versions of the TPSC approximation for the one-band Hubbard model. We showed that both the TPSC+ and the TPSC+SFM approximations maintain some fundamental properties of TPSC: they satisfy the Pauli principle and the Mermin-Wagner theorem, they predict the pseudogap from antiferromagnetic fluctuations in the weak-interaction regime of the $2D$ Hubbard model. The improvements brought by TPSC+ and TPSC+SFM do not come at a significant computational cost. Moreover, these approximations are valid deep in the renormalized classical regime of the $2D$ Hubbard model where TPSC fails. Both TPSC+ and TPSC+SFM hence extend the domain of validity of TPSC. From a quantitative point of view, we showed that TPSC+SFM leads to accurate values of the spin and charge susceptibilities in the doped $2D$ Hubbard model over a wide range of temperatures and interaction strength. For the same model, TPSC+ gives the most accurate results for the self-energy out of the three TPSC approximations considered here. However, TPSC+ does not satisfy the {\it f} sum rule. Our comparisons to the second-order perturbation theory self-energy illustrate the non-trivial nature of the three TPSC approaches. Finally, our work, in line with previous  benchmark efforts \cite{Vilk:1994,Veilleux:1995,Vilk:1996,Vilk_1997,Moukouri:2000,Kyung_2001,TremblayMancini:2011,Schafer_2021,Martin_2023}, shows that TPSC and its variations are reliable methods to obtain qualitative results for the weak-interaction regime of the Hubbard model.
From our benchmark work, we assess that all three TPSC methods are only valid in the weak to intermediate interaction regime of the Hubbard model and that they cannot capture the physics of the strong interaction regime ($U\geq5$). 
 Except deep in the renormalized classical regime, the quantitative improvements brought about by TPSC+ and TPSC+SFM over TPSC are similar to the ones brought about by the recently developed TPSC+DMFT approach \cite{Martin_2023}. 

\textit{Acknowledgments.} We are grateful to Yan Wang for early collaborations and to Moise Rousseau for help with the convergence algorithm. We are especially grateful to the authors of Ref.~\cite{Simkovic_2021}, F. \v{S}imkovic, R. Rossi and M. Ferrero for sharing the CDet results that we used as benchmarks. We are also grateful to T. Schäfer and the authors of Ref.~\cite{Schafer_2021} for making their benchmarks publicly available. This work has been supported by the Natural Sciences and Engineering Research Council of Canada (NSERC) under grant RGPIN-2019-05312 (A.-M.S. T.), by a Vanier Scholarship (C. G.-N.) from NSERC and by the Canada First Research Excellence Fund. Simulations were performed on computers provided by the Canadian Foundation for Innovation, the Minist\`ere de l'\'Education des Loisirs et du Sport (Qu\'ebec), Calcul Qu\'ebec, and Compute Canada.  

\appendix

\section{Details on the implementation of the TPSC+ algorithm}
\label{app:implementation}
\begin{figure}
    \centering
    \includegraphics[width=1\columnwidth]{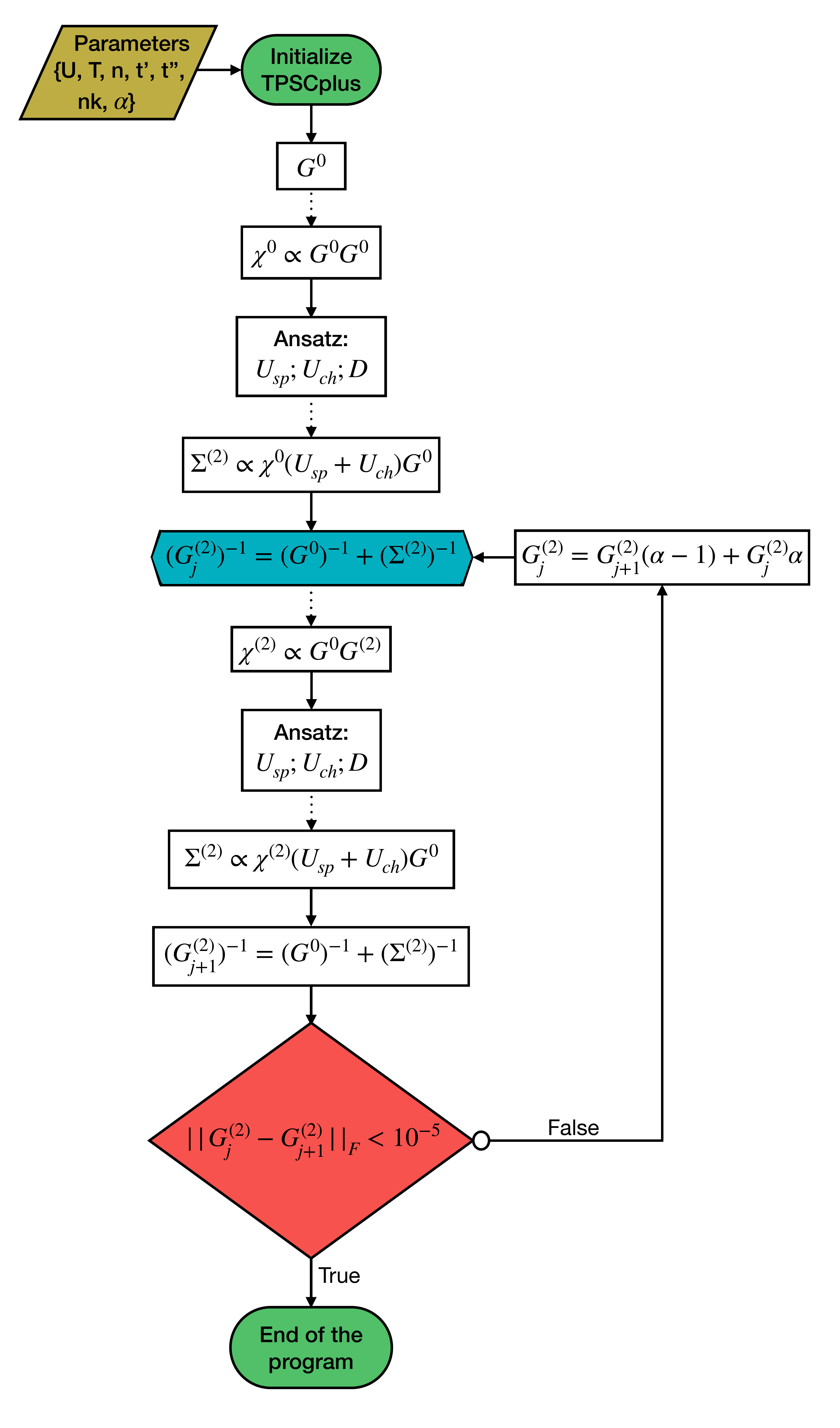}
    \caption{Algorithm's workflow for the TPSC+ method. The input parameters are the Hubbard interaction strenght ($U$), the temperature ($T$), the filling ($n$), the second and third neighbour hopping terms ($t'$,$t''$), and the size of one of the two dimensions in the square reciprocal space ($nk$). The variable $\alpha$ helps convergence by adding damping to the iterative process; it allows to choose the combination of the previous and present Green's function that is input in the new calculation. The dotted lines represents Fast Fourier Transforms. The condition criteria is the Frobenius matrix norm of the difference between the iterated Green's functions at steps $j$ and $j+1$.}
    \label{fig:algo}
\end{figure}

In \fref{fig:algo}, we show the workflow of the TPSC+ calculations. It starts by intitializing the TPSCplus class with the given input parameters. Then begins the TPSC calculation, which we use as the first guess for the self-consistency loop. The implementation for the TPSC+SFM algorithm is similar to the one described in this appendix. 

    
    

The TPSC and TPSC+ approaches allows to choose the size of the reciprocal space that is relevant. We use Fast Fourier transforms (FFT) repetitively, so we choose the number of sites in one spatial direction $n_k$ as a power of 2, because FFT works best in those cases. When  $1/\xi_{sp}$ is smaller than the k-resolution, the results are not considered valid anymore, because in this approximation of TPSC it is important that the self-energy be influenced by long-wavelength spin fluctuations. So the number $n_k$ of the k-resolution should be chosen wisely considering those facts without over-using the resources.

Let the total number of wave vectors be defined as $N=n_k^2$. The correlation functions and the self-energy are defined by convolutions in reciprocal space, which results in order $N^2$ calculations for one physical quantity (susceptibilities or self-energy). It is computationally cheaper to obtain these quantities by Fast Fourier Transforms (FFT). Each FFT takes $N \ln N$ calculations. We use the quantities in the reciprocal space, but because of the convolution properties, it is easy to determine what are those quantities in the real-space. We then have to calculate $F(\Vec{r},\tau) \propto f(\Vec{r},\tau)g(\Vec{r},-\tau)$, instead of $F(\Vec{q},iq_n) \propto \sum_k f(\Vec{k},ik_n)g(\Vec{k}+\Vec{q},ik_n+iq_n)$. So in total, for one physical quantity, one executes $2N \ln N+N$ calculations instead of $N^2$ calculations.

For the calculations of physical quantities, and for handling the many-body propagators with the transitions from real-space to reciprocal-space with Matsubara frequencies, we use sparse sampling and the IR decomposition from the sparse-ir library~\cite{Shinaoka_2017,Li_2020,Shinaoka_sparse_2022}. 

The convergence criterion, as shown in  Fig.~\ref{fig:algo}, is the Frobenius norm, calculated with the numpy function "numpy.linalg.norm()", of the difference between the actual and previous calculations of the interacting Green's function $\mathcal{G}^{(2)}$. When it is met, the calculation is ended. But if it is not, the new Green's function is obtained from a percentage of the actual $(j+1)$ and previous $(j)$ Green's function. The variable that represents this percentage is called $\alpha$. It allows a better convergence, for hard cases. Also, we implemented a loop where the temperature drops slowly and where we use the answer from the previous temperature as a first guess for the next TPSC+ calculation at lower temperature, instead of starting from a new TPSC calculation. It has helped convergence at low temperatures, but it is not a drastic improvement. In even harder cases, at half-filling and at low temperature, we used the "Anderson acceleration" method~\cite{Walker_Ni_2011,Bian_Chen_Kelley_2021}. 

We have compared computing times between TPSC and TPSC+ for the parameters of \fref{fig:comp_gukel_v2}.
The computing time for TPSC+ with the size of the system being 256x256 was roughly 9 seconds and the computing time for TPSC with the same system size was roughly 3 seconds on a personal computer. 




\section{Benchmarks in the strong interaction regime}
\label{app:strongU}

In this Appendix, we show results obtained with TPSC, TPSC+ and TPSC+SFM in the strong interaction regime of the $2D$ Hubbard model. By construction, the three TPSC approaches are not intended to be valid in this regime of parameters. This is mainly due to the formulation of the TPSC ansatz \eref{eq:ansatz}, which is constructed through a Hartree-like decoupling. 

\begin{figure}
    \centering
    \includegraphics[width=1\columnwidth]{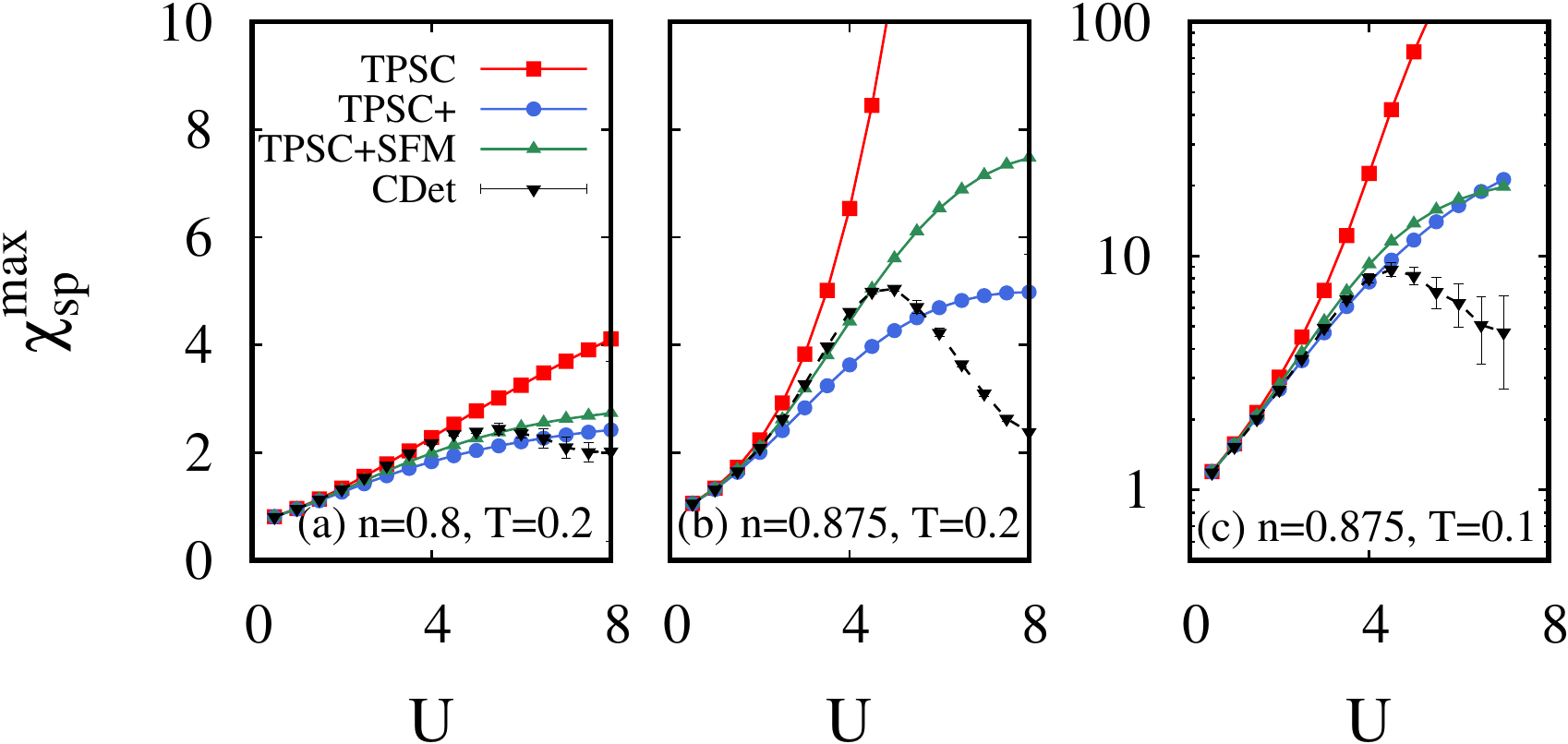}
    \caption{Maximal value of the spin susceptibility obtained from TPSC (red squares), TPSC+ (blue circles) and TPSC+SFM (green triangles) as a function of $U$. Results are shown for (a) $n=0.8$ and $T=0.2$, (b) $n=0.875$ and $T=0.2$, and (c) $n=0.875$ and $T=0.1$. Black lines with error bars are CDet data obtained from Ref. \cite{Simkovic_2021}}
    \label{fig:chisp_strongU}
\end{figure}

We first illustrate the maximal value of the spin susceptibility in \fref{fig:chisp_strongU} as a function of the Hubbard interaction $U$. The model parameters are (a) $n=0.8$ and $T=0.2$, (b) $n=0.875$ and $T=0.2$, and (c) $n=0.875$ and $T=0.1$. The TPSC, TPSC+ and TPSC+SFM results are compared to the CDet benchmark from Ref. \cite{Simkovic_2021}. 
The maximal value of the spin susceptibility increases with $U$ until $U\simeq4$, and decreases as $U$ increases above that point. This maximum near $U=4$ corresponds to the onset of Heisenberg-like physics with a localization of magnetic moments. This behavior is completely missed by the TPSC approach, especially near half filling and at low temperatures: the maximal value of the spin susceptibility only increases with $U$. In contrast, the TPSC+ and TPSC+SFM results do not increase as much with $U$ and even seem to reach a plateau near $U=8$. However, the decrease expected from the CDet data is not seen with these methods above $U=4$, which is a clear indication of the limited interaction regimes that they can reliably access. 

We now consider the following parameter set: density $n=0.8$, temperature $T=0.1$, and Hubbard interaction strength $U=5$. This value of $U$ is slightly above the expected regime of validity of the TPSC methods. We still consider this case in order to study the momentum dependence of the spin and charge susceptibilities, which we can compare here to available CDet data \cite{Simkovic_2021}.

\begin{figure}
    \centering
    \includegraphics[width=1\columnwidth]{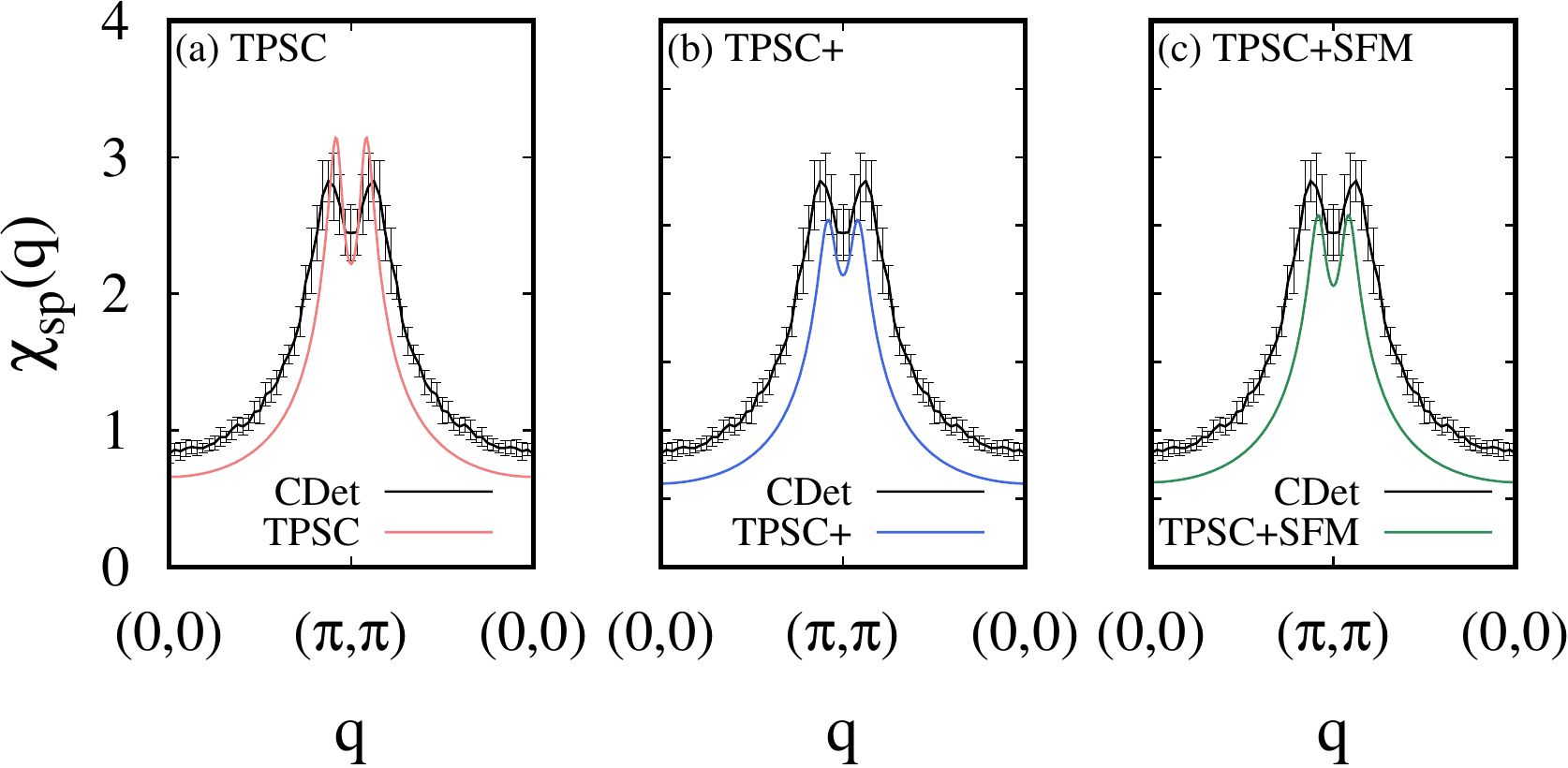}
    \caption{Spin susceptibility along the path (0,0) to ($\pi$,$\pi$) in the Brillouin zone at fixed temperature $T=0.1$, interaction $U=5$ and filling $n=0.8$, obtained with (a) TPSC, (b) TPSC+ and (c) TPSC+SFM calculations. Black lines with error bars are CDet data obtained from Ref. \cite{Simkovic_2021}}
    \label{fig:chisp_along_diag}
\end{figure}

\begin{figure}
    \centering
    \includegraphics[width=1\columnwidth]{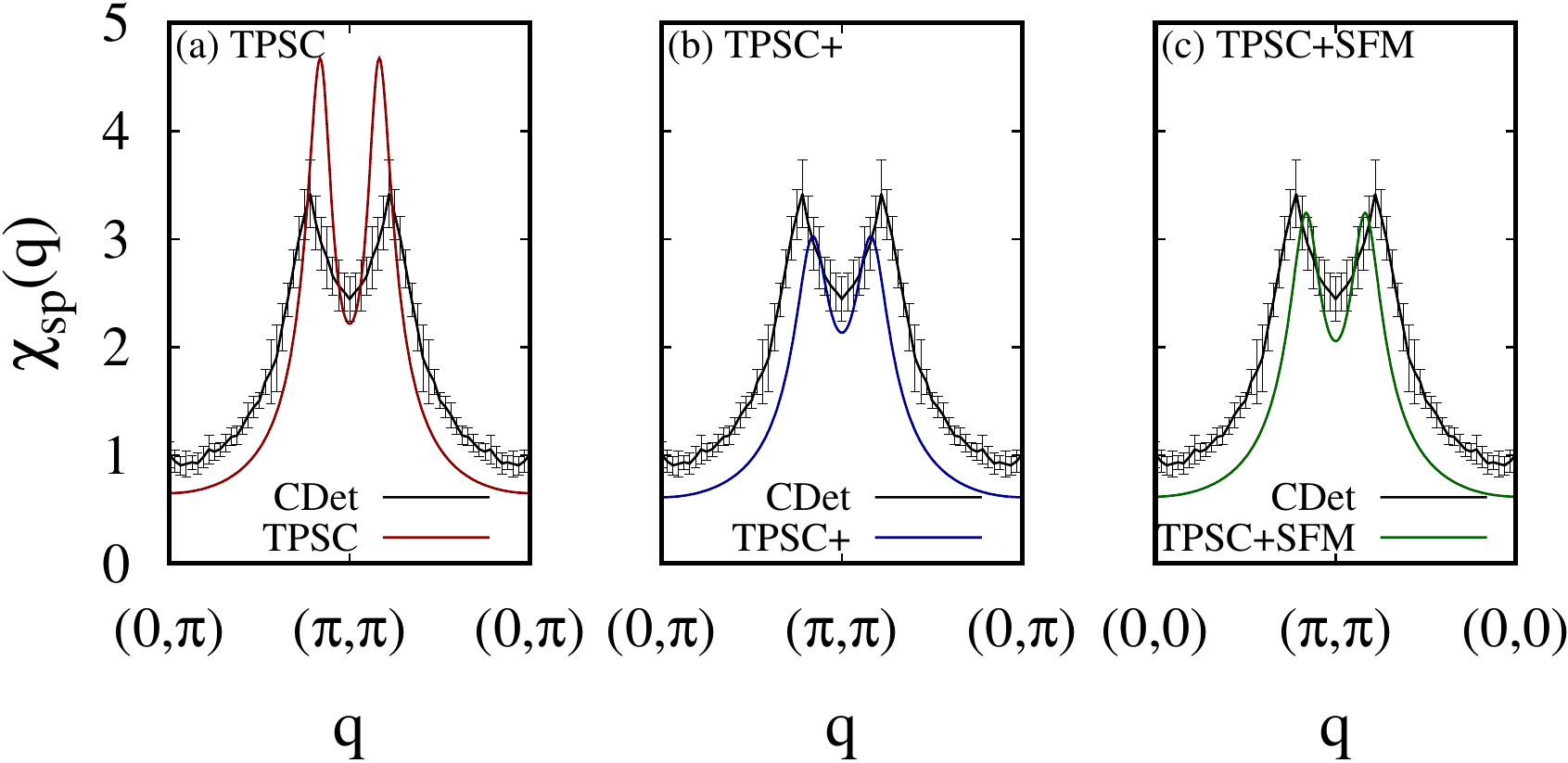}
    \caption{Spin susceptibility along the path (0,$\pi$) to ($\pi$,$\pi$) in the Brillouin zone at fixed temperature $T=0.1$, interaction $U=5$ and filling $n=0.8$, obtained with (a) TPSC, (b) TPSC+ and (c) TPSC+SFM calculations. Black lines with error bars are CDet data obtained from Ref. \cite{Simkovic_2021}}
    \label{fig:chisp_along_edge}
\end{figure}

In \fref{fig:chisp_along_diag} and \fref{fig:chisp_along_edge}, we show the spin susceptibility along the diagonal and along the edge of the Brillouin zone, respectively, for the parameters listed above. Both paths reveal similar information: When compared to CDet, TPSC overestimates the value of the maxima, while TPSC+ and TPSC+SFM underestimate it. The positions of the maxima obtained with the TPSC approaches are slightly shifted with respect to the CDet ones. Still, the results obtained with the three TPSC approaches are in qualitative agreement with the CDet benchmark. The separation of the maximum into two peaks around $(\pi,\pi)$ is captured by these approaches. 


\begin{figure}
    \centering
    \includegraphics[width=1\columnwidth]{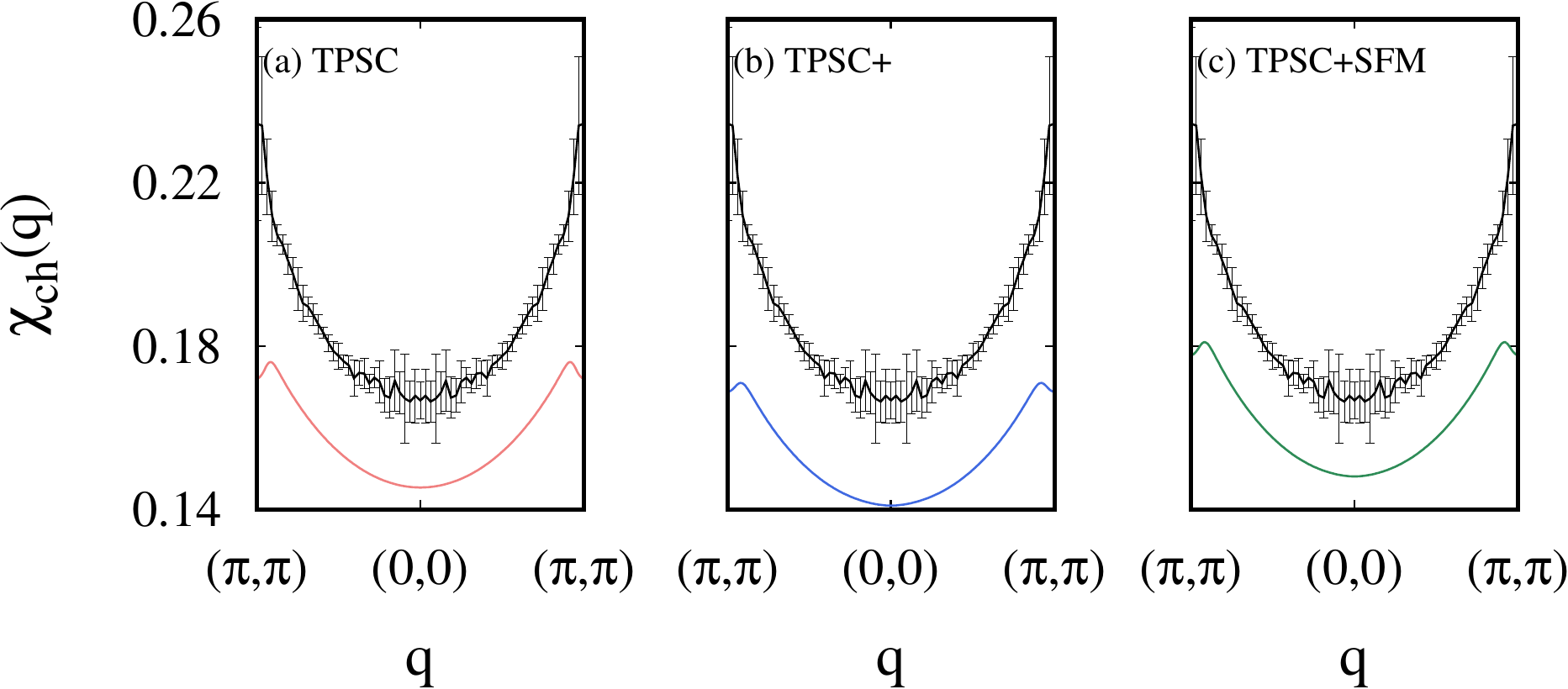}
    \caption{Charge susceptibility along the path ($\pi$,$\pi$) to (0,0) in the Brillouin zone at fixed temperature $T=0.1$, interaction $U=5$ and filling $n=0.8$, from (a) TPSC, (b) TPSC+ and (c) TPSC+SFM calculations. The black lines with error bars are CDet data obtained from Ref. \cite{Simkovic_2021}. }
    \label{fig:path_ch_diag}
\end{figure}

\begin{figure}
    \centering
    \includegraphics[width=1\columnwidth]{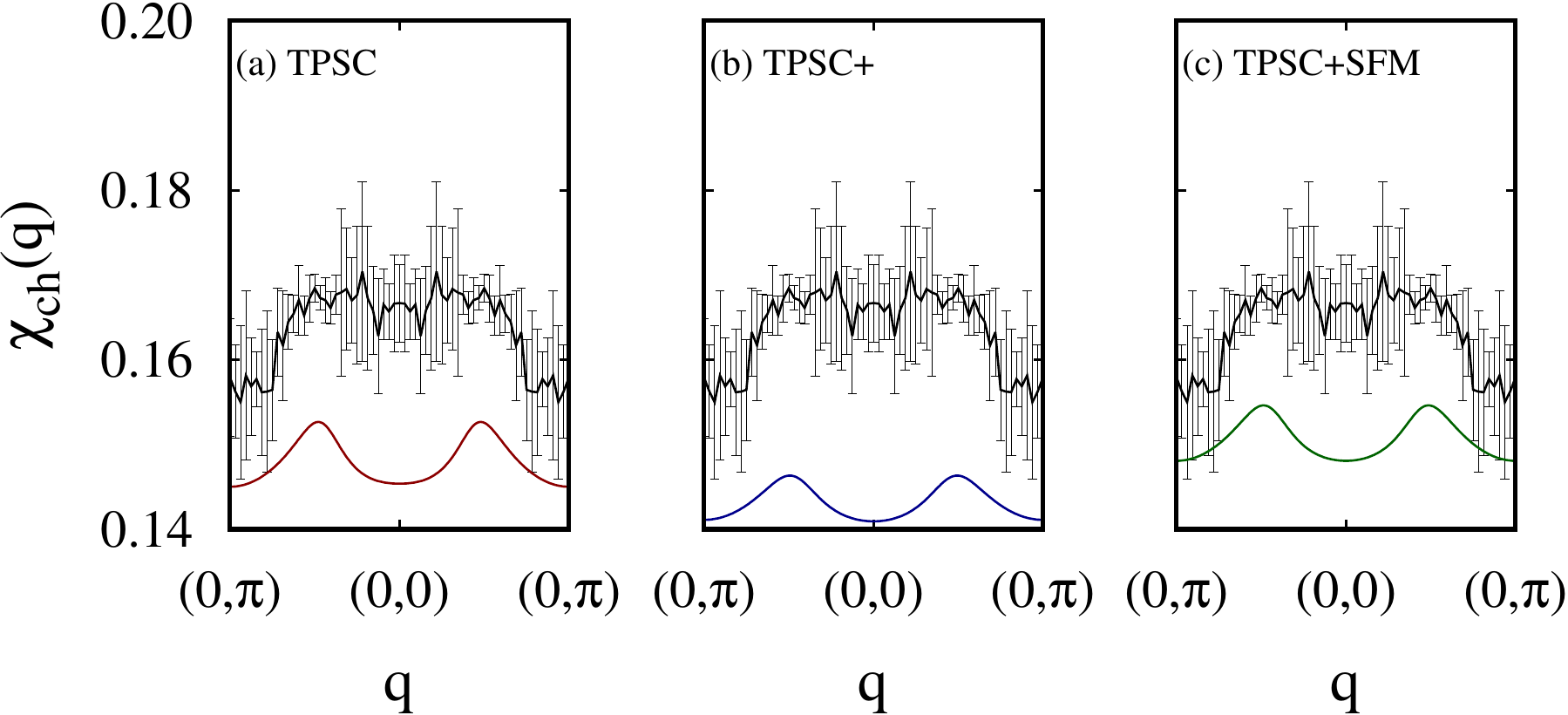}
    \caption{Charge susceptibility along the path (0,$\pi$) to (0,0) in the Brillouin zone at fixed temperature $T=0.1$, interaction $U=5$ and filling $n=0.8$, from (a) TPSC, (b) TPSC+ and (c) TPSC+SFM calculations. The black lines with error bars are CDet data obtained from Ref. \cite{Simkovic_2021}. }
    \label{fig:path_ch_edge}
\end{figure}

In \fref{fig:path_ch_diag} and \fref{fig:path_ch_edge}, we show the charge susceptibility along the edge and the diagonal of the Brillouin zone, respectively. All three TPSC approaches result in lower values of $\chi_{ch}$ than CDet, although they have the right qualitative behavior. As seen in \fref{fig:chichmax_simkovic}, TPSC+SFM is in better agreement with the exact value of $\chi_{ch}$ than TPSC and TPSC+.  

\clearpage


%







    
    


    


\end{document}